\documentclass[10pt,prd,preprintnumbers,floatfix,aps,nofootinbib,showpacs]{revtex4} 

\usepackage{epsfig}
\usepackage{url}
\usepackage{hyperref}
\usepackage[normalem]{ulem} 
\usepackage{latexsym}
\usepackage{epsfig}
\usepackage{amsmath}
\usepackage{amssymb}
\usepackage{graphicx}
\usepackage{verbatim}
\usepackage{enumerate,mdwlist}
\usepackage[titletoc]{appendix}
\usepackage{amsfonts}
\usepackage{tikz} 

\newcommand{\lp}{\left(}
\newcommand{\rp}{\right)}

\newcommand{\ba}{\begin{eqnarray}}
\newcommand{\ea}{\end{eqnarray}}
\newcommand{\be}{\begin{equation}}
\newcommand{\ee}{\end{equation}}

\newcommand{\D}{{\mathcal{D}}}

\newcommand{\M}{\mathcal{M}}
\newcommand{\R}{\mathcal{R}}
\newcommand{\Lag}{\mathcal{L}}

\definecolor{grey}{rgb}{0.4,0.4,0.4}
\definecolor{dullmagenta}{rgb}{0.4,0,0.4}
\definecolor{darkblue}{rgb}{0,0,0.4}
\definecolor{midblue}{rgb}{0,0,0.5}
\definecolor{midred}{rgb}{0.5,0,0}
\definecolor{orange}{rgb}{1,0.5,0}
\definecolor{lightbrown}{rgb}{0.75,0.5,0.25}
\definecolor{tan}{cmyk}{0.14,0.42,0.56,0}
\definecolor{djunglegreen}{cmyk}{0.99,0,0.52,0}
\definecolor{lightgreen}{rgb}{0,1,0}
\definecolor{olivegreen}{cmyk}{0.64,0,0.95,0.40}
\definecolor{midgreen}{rgb}{0.0,0.675,0.0}
\definecolor{darkgreen}{rgb}{0,0.5,0}

\usepackage[all]{xy} 
\usepackage{amsfonts}

\begin{document} 

\title{Towards the most general scalar-tensor theories of gravity:\\ A unified approach in the language of differential forms}

\author{Jose Mar\'ia Ezquiaga}
\email{jose.ezquiaga@uam.es}
\affiliation{Instituto de F\'isica Te\'orica UAM/CSIC, Universidad Aut\'onoma de Madrid, \\ C/ Nicol\'as Cabrera 13-15, Cantoblanco, Madrid 28049, Spain}
\author{Juan Garc\'ia-Bellido}
\email{juan.garciabellido@uam.es}
\affiliation{Instituto de F\'isica Te\'orica UAM/CSIC, Universidad Aut\'onoma de Madrid, \\ C/ Nicol\'as Cabrera 13-15, Cantoblanco, Madrid 28049, Spain}
\author{Miguel Zumalac\'arregui}
\email{miguel.zumalacarregui@nordita.org}
\affiliation{Nordita \\ KTH Royal Institute of Technology and Stockholm University \\
Roslagstullsbacken 23, SE-106 91 Stockholm, Sweden}

\begin{abstract}
We use a description based on differential forms to systematically explore the space of scalar-tensor theories of gravity. Within this formalism, we propose a basis for the scalar sector at the lowest order in derivatives of the field and in any number of dimensions. This minimal basis is used to construct a finite and closed set of Lagrangians describing general scalar-tensor theories invariant under Local Lorentz Transformations in a pseudo-Riemannian manifold, which contains ten physically distinct elements in four spacetime dimensions. Subsequently, we compute their corresponding equations of motion and find which combinations are at most second order in derivatives in four as well as arbitrary number of dimensions. By studying the possible exact forms (total derivatives) and algebraic relations between the basis components, we discover that there are only four Lagrangian combinations producing second order equations, which can be associated with Horndeski's theory. In this process, we identify a new second order Lagrangian, named kinetic Gauss-Bonnet, that was not previously considered in the literature. However, we show that its dynamics is already contained in Horndeski's theory. Finally, we provide a full classification of the relations between different second order theories. This allows us to clarify, for instance, the connection between different covariantizations of Galileons theory. In conclusion, our formulation affords great computational simplicity with a systematic structure. As a first step we focus on theories with second order equations of motion. However, this new formalism aims to facilitate advances towards unveiling the most general scalar-tensor theories.
\end{abstract}

\date{\today}


\maketitle

\section{Introduction}
\label{sec:Introduction}

Gravity is central to many of the unsolved problems in physics, from the origin of the Universe and its fate, to the unification of the fundamental interactions. Despite its fantastic successes, Einstein's theory might not be the final answer and it is necessary to explore different paradigms to shed light on these deep questions. In this sense, alternative theories of gravity can be viewed as effective descriptions of the underlying theory of quantum gravity or tools to solve other theoretical issues, such as the cosmological constant problem.

Recent advances in cosmology also motivate the proposal and investigations of alternatives to Einstein's theory. The discovery of the current era of accelerated expansion \cite{Perlmutter:1998np,Riess:1998cb} requires a radical change in our description of gravity: either by the inclusion of new gravitational degrees of freedom or by the introduction of a tiny cosmological constant that challenges our interpretation of gravity as an effective field theory \cite{Weinberg:1988cp}. 
Moreover, mounting evidence indicates that the early Universe underwent another phase of accelerated expansion, cosmic inflation, that shaped the large scale features of the Universe and seeded perturbations that evolved into galaxies and other large scale structures \cite{Linde:2014nna,Martin:2015dha}. Cosmic inflation could not be caused by a cosmological constant and requires additional degrees of freedom able to strongly affect the gravitational dynamics.

Finally, alternative paradigms are necessary to put our notions of gravity to the test in disparate regimes and honor the effort of experimental collaborations. Earth experiments and Solar System measurements provide very precise data through a variety of post-newtonian effects \cite{Will:2005va}. Cosmological observations of the expansion of the Universe and the evolution of large scale structure provide complementary information on the largest scales available to observation \cite{Clifton:2011jh,Weinberg:2012es,Koyama:2015vza}. Finally, astrophysical systems \cite{Berti:2015itd} such as binary pulsars \cite{Kramer:2009zza} and our central black hole \cite{Psaltis:2008bb} can be used to explore gravity in the strong field regime in which general-relativistic effects are dominant. The recent discovery of gravitational waves from a black hole merger at cosmological distance \cite{Abbott:2016blz} provides a double-edged tool for this effort, allowing us to extract information both from the strong field regime and from the cosmological expansion.

The theoretical questions and the experimental enterprise have motivated the construction of novel, alternative theories of gravity. Among them, scalar-tensor (ST) theories provide the minimal extension of Einstein's theory, with one single additional degree of freedom. Such a degree of freedom, the scalar field, has historically been used in effective field theories to describe phenomena whose energy scale is not accessible, e.g. in the Landau-Ginzburg theory of superconductivity \cite{Ginzburg:1950sr} before Bardeen-Cooper-Schriffer electron-hole pairs and condensate \cite{Bardeen:1957mv}, or the description of pions \cite{Yukawa:1935xg} before the discovery of quarks \cite{Friedman:1972sy}. There could be fundamental scalars like the Higgs, recently discovered at the LHC \cite{Aad:2012tfa,Chatrchyan:2012xdj}, but most of these fields are effective descriptions of a more complicated underlying dynamics, like the scalaron in the case of Starobinsky inflation \cite{Starobinsky:1980te,Mukhanov:1981xt}. In any case, the inclusion of a scalar partner of the graviton in scalar-tensor theories seems the most economical extension of Einstein gravity.

In addition, the simplicity of the scalar field under Lorentz transformations enables one to couple it to the metric in many different ways, allowing the introduction of a rich pattern of possible interactions with the tensor degrees of freedom. Moreover, it is easy to propose ST models with interesting cosmologies, as shown by the plethora of inflationary models considered in the literature \cite{Linde:2014nna,Martin:2015dha}. 
In contrast, other theories with more degrees of freedom are far more restricted. Such is the case of theories with massive gravitons, which were only recently developed \cite{deRham:2010kj,Hassan:2011hr} (see \cite{Hinterbichler:2011tt,deRham:2014zqa} for reviews) and lead to either non-dynamical solutions \cite{DAmico:2011jj}, instabilities \cite{Comelli:2012db} or lack of distinctive signatures \cite{Akrami:2015qga} in their application to cosmology. For this reasons, ST theories have become the standard for tests of gravity as well as models for cosmic acceleration.

In the pursuit of generality, systematic approaches are essential to characterize alternative paradigms and ensure that every possibility is addressed. In this sense, Ostrogradski's theorem allows us to distinguish theories with additional and ghost degrees of freedom caused by higher derivatives in the action \cite{ostrogradski1850member} (for a modern presentation see \cite{Woodard:2015zca}). Furthermore, this result also allows us to classify ST theories depending on the mechanism by which they avoid Ostrogradski's result. The first generation of theories contains no second derivatives of the scalar and are given by generalizations of Jordan-Brans-Dicke theories \cite{Brans:1961sx}. Theories in the second generation are described by second order equations of motion and are characterized by Horndeski's theory \cite{Horndeski:1974wa}. Finally, a third generation of theories with higher-derivative dynamical equations but no additional degrees of freedom has been recently identified \cite{Zumalacarregui:2013pma,Gleyzes:2014dya}. This new family of ST theories is now an active area of research aimed at finding the most general framework for ghost-free ST gravity\footnote{Out of this classification, an alternative route to avoid Ostrogradski's instabilities is to have non-local, infinite derivatives theories \cite{Biswas:2011ar}.}.

A general and systematized classification of gravitational theories is a very challenging task and several attempts have relied on simplifying assumptions in order to construct the most possible general interactions. This has been particularly fruitful in the context of cosmology, where the high degree of symmetry of the background solution facilitates the characterization of possible gravitational interactions order-by-order in the perturbations. This effort led to the effective field theory of inflation \cite{Cheung:2007st}, which was latter generalized to its dark energy analog to explore late-time cosmic acceleration \cite{Gubitosi:2012hu,Bloomfield:2012ff} (with refinements within specific frameworks \cite{Gleyzes:2013ooa,Bellini:2014fua,Zuma:2016tba} and extensions \cite{Lagos:2016wyv}). One of our objectives is to provide the tools to systematically explore this landscape of theories and understand their features without relying on such simplifying assumptions.

At the same time, systematic approaches have also appeared for gravitational theories with only tensorial degrees of freedom. In this field, the basic work was made by Lovelock \cite{Lovelock:1971yv}, who found the most general second order Euler-Lagrange equations for a single massless spin-2 particle in arbitrary dimensions. Then, he found the associated Lagrangian, which is the natural generalization of Einstein's theory. In this sense, Horndeski's theory is just the scalar-tensor extension of Lovelock's theory in four dimensions. To apply these theories to the real world, one must remember that in order to couple fermions to gravity, the gravitational theory must be reformulated in the tangent space \cite{Weyl:1929fm}, which can be easily done using differential forms language \cite{Hehl:1976kj}. From this point of view, systematic studies have been performed too, for instance in Ref. \cite{Zumino:1985dp,Mardones:1990qc}. Despite the actual need of coupling fermions to gravity, the differential form version of Lovelock's theory has been very useful to simplify the computations and understand the inner structure of the theory. Differential forms have also been used in theories involving massive gravitons \cite{Hinterbichler:2012cn}, but such an analysis had not been performed yet in the case of scalar-tensor theories.

In this paper, we are going to investigate the space of ST theories using the language of differential forms. The advantage will be that we are going to find a \emph{finite} and \emph{closed} basis of Lagrangians. Moreover, the antisymmetric structures used to derive, for instance, Horndeski's theory \cite{Horndeski:1974wa} or Generalized Galileons ($G^{2}$) \cite{Deffayet:2009wt}, which could seem ad-hoc at first sight, naturally arise from the requirement that the building blocks of the Lagrangian are differential forms. Remarkably, this approach also clearly disentangles the internal relations between different ST theories, presenting in a \emph{systematic} way all the equivalences through total derivatives or algebraic identities.

In Sec. \ref{sec:RewritingST} we present the set of differential forms that will act as building blocks for our basis of ST theories. Since this section is going to be discussed using the mathematics of differential forms, we have included a summary of the key concepts in Appendix \ref{app:Notation}. In Sec. \ref{sec:EOM} we analyze which Lagrangians of our basis (or combinations thereof) give rise to second order equations of motion, thus becoming automatically free of Ostrogradski's instabilities\footnote{Here and throughout the text, we refer to theories with \emph{covariant} second order equations of motion. Whenever this condition is relaxed, subtleties can arise since, as it was shown in Ref. \cite{Deffayet:2015qwa}, any linear combination of Galileons' Lagrangians can be rewritten in a way in which the equations of motion are second order with respect to time but higher order in space derivatives. However, not all such models are viable as it can be proved analyzing the primary constraints arising from the degeneracy of these Lagrangians \cite{Langlois:2015cwa,Crisostomi:2016czh}.}. We will first consider the scalar equations of motion, Sec. \ref{subsec:sEOM}, and then the tensorial ones, Sec. \ref{subsec:vEOM}. Subsequently, we will study the relations between different second order theories in Sec. \ref{sec:Completeness}. This will allow us to identify which second order Lagrangians are independent. Finally, in Sec. \ref{sec:Discussion}, we will conclude by summarizing the main results and discussing the advantages and potential of our approach.
\section{A General Basis for Scalar-Tensor Theories}
\label{sec:RewritingST}

Scalar-tensor theories are generally described by an action functional $S$, which corresponds to the integral of the Lagrangian $\Lag$ over the curved space-time. In this paper, we are going to exploit the fact that, mathematically, integration is an operation defined in terms of the space of differential forms $\Omega^{q}(\M)$, where $q$ is the order of the $q$-form and the dimension of the base manifold $\M$. Since the action is defined as an integral over a $D$-dimensional curved space-time manifold, the Lagrangian must be a $D$-form, i.e.
\begin{equation}
S=\int_{\M}\Lag.
\end{equation}
Crucially, a $D$-form is characterized for being proportional to the volume element $\eta=\sqrt{-g}dx^{1}\wedge\cdots\wedge dx^{D}$, leading to a direct connection with the usual component notation. Furthermore, due to the fact that $\Omega^{q}(\M)$ is constructed as the space of totally antisymmetric $(0,q)$-tensors, if we construct our $D$-form Lagrangians with exterior products of differential forms, the set of possibilities will be \emph{finite}. 

In order to determine a general basis for scalar-tensor Lagrangians, we must first identify the appropriate building blocks written in differential form language. From the tensorial side, we have the usual geometrical quantities characterizing a manifold. In particular, we will work with differentiable manifolds with an associated metric $\mathfrak{g}$ and 1-form connection $\omega^{a}_{~b}$. Also, we will fix the metric to have a Lorentzian signature. Moreover, we will focus on manifolds with a vanishing torsion $T^{a}=0$ and a metric-compatible connection $\omega_{ab}=-\omega_{ba}$, i.e. \emph{pseudo-Riemannian manifolds}. In such a case, the connection is uniquely determined by the non-coordinate basis elements $\theta^{a}$, which can be related to the curved space-time metric via the flat Minkowski metric $\eta_{ab}$, i.e. $\mathfrak{g}=\eta_{ab}\theta^{a}\otimes\theta^{b}$. Introducing an exterior covariant derivative $\D$ constructed from $\omega^{a}_{~b}$, the geometry of the manifold is encoded in the 2-form curvature, defined as $\R^{a}_{~b}=\D\omega^{a}_{~b}$. This will be our building block characterizing the tensorial part of the action. In components, it reads
\begin{align}
&\R^{a}_{~b}=\frac{1}{2}R^{a}_{~bcd}\theta^{c}\wedge\theta^{d}, \label{eq:Curvature2Form}
\end{align}
where $R^{a}_{~bcd}$ is the corresponding Riemann tensor. One should notice that, throughout the text, we will use latin indices to denote non-coordinate components and greek indices for coordinate ones. Both basis are linked with the vielbein $e^{a}_{~\mu}$ by $\theta^{a}=e^{a}_{~\mu}dx^{\mu}$. Moreover, the 1-form connection $\omega$ and the Levi-Civita connection $\Gamma$ are related by the vielbein postulate $\nabla_{\mu}e^{a}_{~\nu}=0$. In this language, Bianchi's second identity simply implies that $\D\R^{a}_{~b}=0$. In case the reader is not familiar with this notation, we have included in App. \ref{app:Notation} a short review on differential geometry in differential forms language.

Subsequently, we must encounter possible $q$-forms describing the scalar field and its derivatives. The scalar field $\phi$ itself defines a 0-form. Its partial derivative is also a well-defined 1-form, corresponding to the exterior derivative of the scalar field $d\phi=\nabla_{\mu}\phi dx^{\mu}$. However, it is not trivial to introduce the second covariant derivative of the scalar field $\nabla_{\mu}\nabla_{\nu}\phi$ because it is a symmetric $(0,2)$-tensor. Consequently, we must find an appropriate antisymmetric tensor which encodes the information from the second derivatives. Since the tensor is symmetric, we cannot apply directly an antisymmetric operator, i.e. $\nabla_{[\mu}\nabla_{\nu]}\phi=0$. If we apply an antisymmetric operator to only one of the indices, in order to finally obtain a $q$-form, we will end up with a $D$-form, which is a trivial case since it is already proportional to the volume element. Additionally, using Poincare lemma, the exterior derivative of the gradient field vanishes, i.e. $dd\phi=0$. Moreover, by definition, the wedge product of  $d\phi$ with itself is also zero, i.e. $d\phi\wedge d\phi=0$. This means that using this 1-form we could never construct the kinetic term, because it contains two first derivatives. Clearly, we need more adequate definitions of the $q$-forms representing the first and second derivatives of the scalar field. In the following, we propose a minimal setup, in which derivatives of the field appear in the lowest possible order while fulfilling our requirements. This leads to two derivatives of the scalar in each element of the basis. In App. \ref{app:HigherOrder}, we introduce a non-linear generalization of the scalar-tensor theories we are going to present next.

Let us define two vector-valued 1-forms that encode the first and second covariant derivatives of $\phi$
\begin{align}
&\Psi^{a}\equiv\nabla^{a}\phi\nabla_{b}\phi~\theta^{b},  \label{eq:1deriv} \\
&\Phi^{a}\equiv\nabla^{a}\nabla_{b}\phi~\theta^{b}. \label{eq:2deriv} 
\end{align}
Then, we will construct the most general scalar-tensor theory obeying the following:
\begin{itemize}
 \item It is described by an action principle in which the Lagrangian is a $D$-form invariant under Local Lorentz Transformations (LLT) defined in a pseudo-Riemannian manifold.
 \item The Lagrangian is built up out of exterior products of the vielbein $\theta^{a}$, the 2-form curvature $\R^{ab}$, first derivatives of the scalar field $\Psi^{a}$ and second derivatives of the scalar field $\Phi^{a}$. 
\end{itemize}
As a consequence, in order to have a Lagrangian invariant under LLT, there cannot be free indices. Thus, they must be contracted with the tangent space metric $\eta_{ab}$ and the totally antisymmetric symbol $\epsilon_{a_{1}\cdots a_{D}}$, which are invariant objects\footnote{We will choose the convention $\eta_{ab}=\mathrm{diag}(-1,1,1,1)$ for the metric signatures and $\epsilon_{0123}=+1$ for the antisymmetrizations.}. Moreover, the fact that we restrict to pseudo-Riemannian manifolds, i.e. manifolds with a metric-compatible connection and a vanishing torsion, implies that all the tensorial dynamics is contained in the 2-form curvature (\ref{eq:Curvature2Form}). With these two conditions, we can define a basis of Lagrangian given by
\begin{equation}
\Lag_{(lmn)}=\bigwedge_{i=1}^{l}\mathcal{R}^{a_{i}b_{i}}\wedge\bigwedge_{j=1}^{m}\Phi^{c_{j}}\wedge\bigwedge_{k=1}^{n}\Psi^{d_{k}}\wedge\theta^{\star}_{~a_{1}b_{1}\cdots a_{l}b_{l}c_{1}\cdots c_{m}d_{1}\cdots d_{n}},
\label{eq:L}
\end{equation}
where $\bigwedge$ is an abbreviation for a set of consecutive wedge products and $l,m,n\in\mathbb{N}$. In this notation, if any of the subindices of the Lagrangian are zero, the corresponding terms in the r.h.s do not appear. Here, $\theta^{\star}_{a_{1}\cdots a_{k}}$ is the Hodge dual basis and it is defined as 
\begin{equation}
\theta^{\star}_{~a_{1}\cdots a_{k}}=\frac{1}{(D-k)!}\epsilon_{a_{1}\cdots a_{k}a_{k+1}\cdots a_{D}}\theta^{a_{k+1}}\wedge\cdots\wedge\theta^{a_{D}}.
\label{eq:HD}
\end{equation}
One should notice that the previous result $d\phi\wedge d\phi=0$ appears in this notation making $\mathcal{L}_{(lmn)}$ vanish for $n>1$. Additionally, it must hold that $2l+m+n\leq D$ due to the antisymmetry by the Hodge dual basis. This will be very important because it means that for a given dimension $D$ our basis of Lagrangians will be finite. Interestingly, if we do not include the scalar field, setting $m=n=0$, these Lagrangians correspond to Lovelock's theory \cite{Lovelock:1971yv} written in differential forms (see \cite{Charmousis:2008kc} for a modern summary using our notation). Therefore, this basis of Lagrangians could be seen as its \emph{scalar-tensor extension}. Finally, it is important to remark that there are three additional Lagrangians that fulfill our premises but are not included in our basis (\ref{eq:L}). They correspond to Lagrangians in which the indices of the building blocks are contracted among them, e.g. $\R^{ab}\wedge\Phi_{a}\wedge\Psi_{b}$. However, they do not lead to second order equations of motion. Thus, we discard them from the beginning. For completeness, we present them in App. \ref{app:PontryaginForms}.

In the scalar-tensor theories represented by the basis (\ref{eq:L}), the action will be the sum over all possible Lagrangians with different $l$, $m$ and $n$ integrated over the space-time manifold, i.e.
\begin{equation}
S=\sum_{l,m,n}^{p\leq D}\int_{\mathcal M}\alpha_{lmn}\mathcal{L}_{(lmn)},
\label{eq:Action}
\end{equation}
where $p\equiv2l+m+n$ and $n\leq 1$. In this context, the coefficients $\alpha_{lmn}$ represent 0-forms, which, in general\footnote{Here, it will be important that the coefficient is a 0-form and that we are constructing the geometrical quantities out of the 2-form curvature $\R^{ab}$. Consequently, we will not consider any dependence in curvature scalars in $\alpha_{lmn}$, e.g. $R^{2}$ or $R_{ab}R^{ab}$. In this sense, we will not be covering theories such as $f(R)$ \cite{Sotiriou:2008rp} or more generally $f({\rm Lovelock})$ \cite{Bueno:2016dol}, which are automatically free of Ostrogradski's instabilities. Nevertheless, such theories can be described as scalar-tensor theories in most cases \cite{Bueno:2016dol}.}, can be functions of the scalar field and its derivatives $\alpha_{lmn}=\alpha_{lmn}(\phi,X,[\Phi],\cdots)$, where we are using the notation, exemplified in detail in App. \ref{app:Contractions}, for which a square bracket represents the contraction of two free indices, e.g. $[t_{\mu\nu}]\equiv t^{\mu}_{~\mu}$, and an angle bracket the contraction with partial derivatives of the scalar field, e.g. $\langle t_{\mu\nu}\rangle\equiv\phi^{,\mu}t_{\mu\nu}\phi^{,\nu}$. Also, partial derivatives are shortened by a comma, $\partial_{\mu}\phi=\phi_{,\mu}$, and covariant derivatives are shortened by a semicolon, $\nabla_{\mu}\nabla_{\nu}\phi=\phi_{;\mu\nu}$. Lastly, we write the contractions of second derivatives as $\left.\Phi^{n}\right._{\mu\nu}=\phi_{;\mu\alpha_{1}}\left.\phi^{;\alpha_{1}}\right._{;\alpha_{2}}\cdots\left.\phi^{;\alpha_{n-1}}\right._{;\nu}$ and define $-2X\equiv \phi^{,\mu}\phi_{,\mu}$.

In $4D$, we have 15 possible Lagrangians in our basis. In order to translate them into the usual component notation, we only need to apply the definition of the wedge product and the Hodge dual basis. For completeness, we present in App. \ref{app:Contractions} the explicit component expression for a general $\Lag_{(lmn)}$. Here, we show for the first cases how this general recipe works. Recalling that $\eta=\theta^{1}\wedge\cdots\wedge\theta^{D}$ is the volume element, we find the following Lagrangians:
\begin{enumerate}[(i)] \itemsep0pt \parskip0pt \parsep0pt
\item $p=0$
\begin{align}
\mathcal{L}_{(000)}&=\theta^{\star}=\eta, \label{eq:L000} 
\intertext{\item $p=1$}
\mathcal{L}_{(010)}&=\Phi^{a}\wedge\theta^{\star}_{~a}=\frac{1}{3!}\phi^{;a}_{~~;e}\epsilon_{abcd}\epsilon^{ebcd}\eta=[\Phi]\cdot\eta, \label{eq:L010} \\
\mathcal{L}_{(001)}&=\Psi^{a}\wedge\theta^{\star}_{~a}=\frac{1}{3!}\phi^{,a}\phi_{,e}\epsilon_{abcd}\epsilon^{ebcd}\eta=-2X\cdot\eta, \label{eq:L001} 
\intertext{\item $p=2$}
\mathcal{L}_{(100)}&=\R^{ab}\wedge\theta^{\star}_{~ab}=\frac{1}{2\cdot2!}R^{a}_{~bef}\epsilon_{abcd}\epsilon^{efcd}\eta=R\cdot\eta, \label{eq:L100} \\
\mathcal{L}_{(020)}&=\Phi^{a}\wedge\Phi^{b}\wedge\theta^{\star}_{~ab}=\frac{1}{2!}\phi^{;a}_{~~;e}\phi^{;b}_{~~;f}\epsilon_{abcd}\epsilon^{efcd}\eta=([\Phi]^2 -[\Phi^2])\eta, \label{eq:L020} \\
\mathcal{L}_{(011)}&=\Phi^{a}\wedge\Psi^{b}\wedge\theta^{\star}_{~ab}=\frac{1}{2!}\phi^{;a}_{~~;e}\phi^{,b}\phi_{,f}\epsilon_{abcd}\epsilon^{efcd}\eta=-(\langle\Phi\rangle+2X[\Phi])\eta, \label{eq:L011} 
\intertext{\item $p=3$}
\mathcal{L}_{(110)}&=\mathcal{R}^{ab}\wedge\Phi^{c}\wedge\theta^{\star}_{~abc}=-2G^{ab}\Phi_{ab}\eta, \label{eq:L110} \\
\mathcal{L}_{(030)}&=\Phi^{a}\wedge\Phi^{b}\wedge\Phi^{c}\wedge\theta^{\star}_{~abc}=([\Phi]^3 -3[\Phi][\Phi^2]+2[\Phi^3])\eta, \label{eq:L030} \\
\mathcal{L}_{(101)}&=\mathcal{R}^{ab}\wedge\Psi^{c}\wedge\theta^{\star}_{~abc}=-2\langle G\rangle\eta, \label{eq:L101} \\
\mathcal{L}_{(021)}&=\Phi^{a}\wedge\Phi^{b}\wedge\Psi^{c}\wedge\theta^{\star}_{~abc}=2(\langle\Phi^2\rangle-\langle\Phi\rangle[\Phi]-X([\Phi]^{2}-[\Phi^2]))\eta, \label{eq:L021} 
\intertext{\item $p=4$}
\mathcal{L}_{(200)}&=\mathcal{R}^{ab}\wedge\mathcal{R}^{cd}\wedge\theta^{\star}_{~abcd}=(R_{abcd}R^{abcd}-4R_{ef}R^{ef}+R^{2})\eta, \label{eq:L200} \\
\mathcal{L}_{(120)}&=\mathcal{R}^{ab}\wedge\Phi^{c}\wedge\Phi^{d}\wedge\theta^{\star}_{~abcd}=(R([\Phi]^{2}-[\Phi^2])-4R^{ab}([\Phi]\Phi_{ab}-\Phi^{2}_{ab})+2R^{abcd}\Phi_{ac}\Phi_{bd})\eta, \label{eq:L120} \\
\mathcal{L}_{(040)}&=\Phi^{a}\wedge\Phi^{b}\wedge\Phi^{c}\wedge\Phi^{d}\wedge\theta^{\star}_{~abcd}=([\Phi]^4 -6[\Phi]^2[\Phi^2]+3[\Phi^2]^2+8[\Phi][\Phi^3]-6[\Phi^4])\eta, \label{eq:L040} \\
\mathcal{L}_{(111)}&=\mathcal{R}^{ab}\wedge\Phi^{c}\wedge\Psi^{d}\wedge\theta^{\star}_{~abcd} \label{eq:L111} \\
&=\left(4\left(\langle R_{ab}\Phi^{bc}\rangle+X[R\Phi]\right)-R\left(\langle\Phi\rangle+2X[\Phi]\right)+2\left(\langle R_{abcd}\Phi^{bd}\rangle-\langle R\rangle[\Phi]\right)\right)\eta, \nonumber\\
\mathcal{L}_{(031)}&=\Phi^{a}\wedge\Phi^{b}\wedge\Phi^{c}\wedge\Psi^{d}\wedge\theta^{\star}_{~abcd}  \label{eq:L031} \\
&=(6(\langle\Phi^2\rangle[\Phi]-\langle\Phi^3\rangle)-3\langle\Phi\rangle([\Phi]^{2}-[\Phi^2])-2X([\Phi]^{3}-3[\Phi][\Phi^{2}]+2[\Phi^{3}]))\eta, \nonumber
\end{align}
\end{enumerate}
where $R$ is the Ricci scalar, $R_{ab}$ is the Ricci tensor and $G_{ab}$ is the Einstein tensor, given by $G_{ab}=R_{ab}-\frac{1}{2}g_{ab}R$. As a consequence of the above expressions, we can easily relate our results with the current literature. For instance, the modern version of Horndeski's Theory \cite{Horndeski:1974wa} is a linear combination of (\ref{eq:L000}), (\ref{eq:L010}), (\ref{eq:L100}-\ref{eq:L020}) and (\ref{eq:L110}-\ref{eq:L030}), and the class of viable theories Beyond Horndeski known as Generalized Generalized Galileons ($G^3$) \cite{Gleyzes:2014dya} are simply (\ref{eq:L021}) and (\ref{eq:L031}). In addition, terms such as (\ref{eq:L101}) and (\ref{eq:L111}) appear when doing a Kaluza-Klein compactification of higher dimensional Lovelock's densities \cite{VanAcoleyen:2011mj} and correspond respectively to ``John" and ``Paul" Lagrangians of the Fab Four theory \cite{Charmousis:2011bf}. Furthermore, when we are in flat space, Galileon theory \cite{Nicolis:2008in} is built up with (\ref{eq:L010}), (\ref{eq:L001}), (\ref{eq:L011}), (\ref{eq:L021}) and (\ref{eq:L031}). In this work, we will show that there is a well-established interconnection between all these Lagrangians. In fact, not all of them are independent, as we will see in Sec. \ref{sec:Completeness}, and only certain linear combinations give rise to second order equations of motion, cf. Sec. \ref{sec:EOM}.

Before computing the Euler-Lagrange equations, we should consider an extension of our basis (\ref{eq:L}), which naturally appears when one applies an exterior derivative to the previous expressions. Acting $\D$ on the scalar one-forms $\Psi^{a}$ and $\Phi^{a}$, given in (\ref{eq:1deriv}) and (\ref{eq:2deriv}), we find\footnote{In components, they will read $\D\Psi^{a}=\nabla_{[b}\left(\nabla^{a}\phi\nabla_{c]}\phi\right)\theta^{b}\wedge\theta^{c}=\nabla_{[b}\left(\nabla^{a}\phi\right)\nabla_{c]}\phi\theta^{b}\wedge\theta^{c}=\Phi^{a}\wedge d\phi$ and $\D\Phi^{a}=\nabla_{[b}\left(\nabla^{a}\nabla_{c]}\phi\right)\theta^{b}\wedge\theta^{c}=\frac{1}{2}R^{a}_{~dbc}\nabla^{d}\phi\theta^{b}\wedge\theta^{c}=\R^{a}_{~d}\nabla^{d}\phi$.}
\begin{align}
\D\Psi^{a}=&\D(\nabla^{a}\phi)\wedge \D\phi+\nabla^{a}\phi\wedge \D(\D\phi)=\Phi^{a}\wedge \D\phi,  \label{eq:d1deriv} \\
\D\Phi^{a}=&\D(\D(\nabla^{a}\phi))=d\omega^{a}_{~b}\wedge\phi^{,b}+\omega^{a}_{~c}\wedge\omega^{c}_{~b}\wedge\phi^{,b}=\mathcal{R}^{a}_{~z}\nabla^{z}\phi, \label{eq:d2deriv} 
\end{align}
where we have used the explicit definition of the covariant derivative $\D v^{a}=dv^{a}+\omega^{a}_{~b}\wedge v^{b}$ and the fact that $d\phi=\D\phi$. Also, we have assumed a vanishing torsion, which in terms of $\D$ reads $T^{a}=\D\theta^{a}=0$, implying that $\D\theta^{\star}_{~a_{1}\cdots a_{k}}=0$. Lastly, we have used that $\Phi^{a}=\D(\nabla^{a}\phi)$. Thus, a covariant exterior derivative applied on $\Phi^{a}$ introduces a contraction of the 2-form curvature with the first derivative of the scalar field. Moreover, if we apply this derivative to a general coefficient $\alpha_{lmn}=\alpha_{lmn}(\phi,X,[\Phi],\cdots)$, we obtain
\begin{align}
\D\alpha_{lmn}=&\alpha_{lmn,\phi}\D\phi-\alpha_{lmn,X}\nabla_{a}\phi\Phi^{a}+\mathcal{O}\left(\alpha_{lmn,[\Phi]}\right),  \label{eq:dalpha}
\end{align}
where $\alpha_{lmn,\phi}=\partial\alpha_{lmn}/\partial\phi$ and $\alpha_{lmn,X}=\partial\alpha_{lmn}/\partial X$. Here, $\mathcal{O}\left(\alpha_{lmn,[\Phi]}\right)$ encodes higher order terms coming from the dependence of $\alpha_{lmn}$ in second order derivative scalars such as $[\Phi]$. Again, we observe that, when we apply exterior derivatives, contractions of the building blocks with gradient fields appear. For that reason, we enlarge the two defining conditions of the basis of Lagrangians $\Lag_{(lmn)}$ presented above to allow contractions with the gradient field $\nabla^{a}\phi$. In the following, we summarize all the relevant new terms of the extended basis in 
\begin{equation}
\Lag_{(\bar{l}m0)}=\nabla_{\bar{a}_{1}}\phi\mathcal{R}^{\bar{a}_{1}b_{1}}\wedge\bigwedge_{i=2}^{l}\mathcal{R}^{a_{i}b_{i}}\wedge\bigwedge_{j=1}^{m}\Phi^{c_{j}}\wedge\theta^{\star}_{~a_{1}b_{1}\cdots a_{l}b_{l}c_{1}\cdots c_{m}}\nabla^{a_{1}}\phi
\label{eq:Lbarl}
\end{equation}
and
\begin{equation}
\Lag_{(l\bar{m}0)}=\bigwedge_{i=1}^{l}\mathcal{R}^{a_{i}b_{i}}\wedge\Phi^{\bar{c}_{1}}\nabla_{\bar{c}_{1}}\phi\wedge\bigwedge_{j=2}^{m}\Phi^{c_{j}}\wedge\theta^{\star}_{~a_{1}b_{1}\cdots a_{l}b_{l}c_{1}\cdots c_{m}}\nabla^{c_{1}}\phi,
\label{eq:Lbarm}
\end{equation}
where we have introduced a bar over the indices of $\Lag_{(lmn)}$ to indicate that a contraction with a gradient field has been performed. Importantly, due to the antisymmetry of the Hodge dual basis, only one element can be contracted at a time. Additionally, a $\Psi^{a}$ term is also incompatible with a contraction since $\Lag_{(\bar{l}m1)}=\Lag_{(l\bar{m}1)}=0$ and $\Lag_{(lm\bar{1})}=-2X\Lag_{(lm1)}$. As before, this general Lagrangian written in differential forms can be translated into components. We include the component expression of the 10 possible contracted Lagrangians in $4D$ in App. \ref{app:Contracted}. 

\section{Equations of Motion}
\label{sec:EOM}

In order to obtain the equations of motion (e.o.m.), we must vary the action (\ref{eq:Action}). When we vary with respect to the frame $\theta^{a}$, we end up with the \emph{vielbein e.o.m} (equivalent to Palatini variation). In this respect, we will restrict our computations to the second order formalism, i.e. we will consider the connection 1-form as a unique function of the vielbein, $\omega=\omega(\theta)$ (equivalent to metric variation). However, as it will become clear later, our approach can be easily extrapolated to a first order formalism, where $\omega^{ab}$ and $\theta^{a}$ are independent variables. When we vary with respect to the scalar field $\phi$, we obtain the \emph{scalar e.o.m}. We will include a 0-form coefficient in front of every Lagrangian, namely $\alpha_{lmn}$. The functional dependence of these coefficients $\alpha_{lmn}=\alpha_{lmn}(\phi,X,[\Phi],\cdots)$ will be relevant to derive the e.o.m. In fact, it will be a crucial ingredient for analyzing the derivative order of the e.o.m. Our goal will be to compute the e.o.m. associated to each possible $\Lag_{(lmn)}$. Then, we will look for combinations of those Lagrangians giving rise to second order e.o.m., thus automatically evading Ostrogradski instabilities \cite{ostrogradski1850member}. Since there is a well defined hierarchy in terms of the number of fields, or the number $p$, defined as $p\equiv2l+m+n$, we will follow that order to obtain the e.o.m., from lower to higher $p$.

Before computing the e.o.m., we will have to work out some expressions. Since we want to work in a manifestly covariant way, we will use the exterior covariant derivative $\D$. In this notation, Cartan's structure equations are simply $T^{a}= \D\theta^{a}$ and $\mathcal R^{a}_{~b}= \D\omega^{a}_{~b}$. Accordingly, Bianchi identities read $\D T^{a}=\mathcal R^{a}_{~b}\wedge\theta^{b}$ and $\D \mathcal R ^{a}_{~b}=0$. For the scalar 1-forms $\Psi^{a}$ and $\Phi^{a}$, we have already seen that acting with $\D$ yields (\ref{eq:d1deriv}) and (\ref{eq:d2deriv}). Also, we will make use of the generalized Stoke's theorem \cite{nakahara2003geometry},
\begin{equation}
\int_{\M}\D\omega=\int_{\partial\mathcal M}\omega.
\label{eq:Stokes}
\end{equation}
Assuming that the boundary contribution vanishes, this could be used, for instance, to move the derivative from one $q$-form to another in a wedge product, recalling that the exterior derivative follows a graded Leibniz rule, i.e. $d(\alpha_{q}\wedge\beta_{r})=(d\alpha_{q})\wedge\beta_{r}+(-1)^{q}\alpha_{q}\wedge(d\beta_{r})$, where $\alpha_{q}$ is a $q$-form. In addition, we will use the following identities
\begin{align}
\nabla^{a}\Phi^{z}&=\nabla^{z}\Phi^{a}-\mathfrak{i}_{\nabla\phi}\R^{az}, \label{eq:id1} \\
\nabla^{z}\R^{ab}\wedge \theta^{\star}_{~ab}&=-2\nabla^{a}\R^{bz}\wedge \theta^{\star}_{~ab}, \label{eq:id2}
\end{align}
where $\mathfrak{i}_{V}\omega$ represents the interior product\footnote{For more details in the definition of this operation, one can read App. \ref{app:Notation}, where the component expression is presented in (\ref{eq:intprod}).} of a $q$-form $\omega$ with respect to a vector field $V$. The first identity, (\ref{eq:id1}), is just the differential form version of the commutator of two covariant derivatives. The second one, (\ref{eq:id2}), is the analog of Bianchi's second identity. Furthermore, thanks to the vanishing of the torsion, $T^{a}=\D\theta^{a}=0$, and the vielbein postulate, $\nabla_{\mu}e^{a}_{~\nu}=0$, the derivatives act trivially on the Hodge dual basis, i.e. $\D\theta^{\star}_{~a_{1}\cdots a_{k}}=0$ and $\nabla^{b}\theta^{\star}_{~a_{1}\cdots a_{k}}=0$. 

Finally, one should notice that the possible higher derivative terms, meaning higher than two derivatives in $\phi$ or $\theta^{a}$, will be $\nabla^{z}\Phi^{a}$ and $\nabla^{z}\R^{ab}$. However, whenever all of the indices are contracted with the Hodge dual basis, the previous terms are not dangerous due to the antisymmetry of its indices. In the first case, a commutator of covariant derivatives naturally appears. In the second case, the whole expression vanishes due to Bianchi second identity. To make this point more visual, we will use the first letters of the latin alphabet, $a,b,c,\cdots$, to indicate indices of the Hodge dual basis. Conversely, we will use the last letters of the latin alphabet, $z,y,x,\cdots$, to denote indices not contracted with the Hodge dual. Indeed, one could notice that we have implicitly introduced this index notation in previous expressions. Furthermore, to make the computations as clear as possible, we will underline the dangerous higher derivative terms. When the line is dashed, e.g. $\dashuline{\nabla^{z}\Phi^{a}}$ \vspace{-4pt}, it will indicate that this particular higher derivative term is compensated with another term in the same expression. This cancellation of higher derivatives will be caused by the commutation of covariant derivatives, (\ref{eq:id1}), or by applying Bianchi identity, (\ref{eq:id2}). On the contrary, when the higher derivative term is not cancelled within that expression, we will underline the term with a solid line, e.g. $\uline{\nabla^{z}\R^{ab}}$. The philosophy will be to investigate if the remaining solid underlined terms of different Lagrangians can be eliminated by choosing appropriate coefficients among them. If the final Lagrangian can be \emph{directly} related with the standard formulation of Horndeski's theory, we will dub it $\Lag^{H}_{i}$. For the rest, we will write $\Lag_{i}^{NH}$. Later on, in Sec. \ref{sec:Completeness}, we will totally clarify the role of $\Lag_{i}^{NH}$ and its connection with $\Lag^{H}_{i}$.

\subsection{Scalar Equations of Motion}
\label{subsec:sEOM}

We begin the computation with the scalar e.o.m. As it was stated before, we will classify the different Lagrangians by the order of the $q$-form constructed with the curvature 2-form and the first and second derivative one-forms, i.e. by an increasing number $p$. At each level, we will consider separately the Lagrangians with $n=0$ and $n=1$, because they have a different structure. One should notice that the only building blocks of our basis depending on $\phi$ are $\Psi^{a}$ and $\Phi^{a}$. Their variations with respect to the scalar field follow
\begin{align}
\delta\Psi^{a}&=\nabla^{a}\delta\phi\D\phi+\nabla^{a}\phi\D\delta\phi, \label{eq:varPsi} \\
\delta\Phi^{a}&=\D\nabla^{a}\delta\phi, \label{eq:varPhi}
\end{align}
where we have used that the variation commutes with the covariant derivatives. Additionally, we will have to consider the variation of the coefficient in front of each Lagrangian $\Lag_{(lmn)}$. In the following, we will focus first in the case of $4D$, where we will denote the coefficients by $G_{i}$, $F_{i}$, $E_{i}$ and $H_{i}$ for shortness. Since the structure of the computation will be very similar, we will only include the details of the calculations for the first cases. For the rest, we will include the full result only in App. \ref{app:sEOM}. Afterwards, we will generalize the result to $D$-dimensions, where we will use the general notation for the coefficients $\alpha_{lmn}$ because it will be important to analyze terms with different $(lmn)$. As we have mentioned above, the coefficients can, in principle, depend on higher derivative scalars, e.g. $[\Phi]$. However, we will argue in the next computation that they must depend only in $\phi$ and $X$ in order to have second-order equations.
\begin{enumerate}[(i)] 
\item $p=0$

For this first case, we only need to consider $G_{2}\Lag_{(000)}=G_{2}\wedge\theta^{\star}$. The scalar e.o.m are simply given by the variation of the coefficient $G_{2}=G_{2}(\phi,X,[\Phi],\cdots)$, i.e.
\begin{equation}
\begin{split}
\delta(G_{2}\Lag_{(000)})=&\delta G_{2}\wedge\theta^{\star} \\
=&\left(\frac{\partial G_{2}}{\partial\phi}\delta\phi+\frac{\partial G_{2}}{\partial\phi^{,z}}\nabla^{z}\delta\phi+\frac{\partial G_{2}}{\partial\phi^{;yz}}\nabla^{y}\nabla^{z}\delta\phi+\cdots\right)\wedge\theta^{\star}.
\end{split}
\end{equation}
Therefore, in order to obtain 2nd order e.o.m, we only need to impose $G_{2}=G_{2}(\phi,X)$. Importantly, this will also happen for the rest of the cases. The point is that if the coefficient depends on second derivatives or higher, there will always be a higher derivative term in the e.o.m. proportional to the original Lagrangian, e.g. $\nabla^{z}\nabla_{z}(G_{i,[\Phi]})\Lag_{(lmn)}$. Thus, it cannot be cancelled with another term in the e.o.m. since any other variation changes the original Lagrangian\footnote{Unless $G_{i,[\Phi]}$ only depends on first derivatives and these are canceled by other terms, cf. (\ref{eq:H1},\ref{eq:H1a}). However, this would only work if $G_{i}$ reproduces the component expression of any of the Lagrangians $\Lag_{(lmn)}$ considered in the next computations. Obviously, this would imply a repetition of the same results. Thus, in the following, we are going to set $G_{i}=G_{i}(\phi,X)$ to eliminate this degeneracy.}. Moreover, it cannot be eliminated with similar terms from other Lagrangians since they will be proportional to different $\Lag_{(lmn)}$. Consequently, we will have to impose $G_{i}=G_{i}(\phi,X)$ in the following calculations. In this particular case, $\Lag_{(000)}=\eta$, choosing $G_{2}=G_{2}(\phi,X)$ also eliminates the possible degeneracy of having $G_{2}$ equal to the component form of any other $\Lag_{(lmn)}$, see (\ref{eq:L010}-\ref{eq:L031}). Nevertheless, there is a remaining degeneracy between $G_{2}(\phi,X)\wedge\Lag_{(000)}$ and $\Lag_{(001)}=-2X\eta$ that we will deal with in Sec. \ref{subsec:AntisymDegeneracies}. In conclusion, for $p=0$, we have found that the e.o.m. of the following Lagrangian are at most 2nd order
\begin{equation}
\Lag_{2}^{H}=G_{2}\wedge\theta^{\star}=G_{2}(\phi,X)\eta.
\label{eq:H2}
\end{equation}
Since this Lagrangian can be directly linked with Horndeski's theory, we have used the superscript $H$.

\item $p=1$

At this level, we have two possible Lagrangians, $\Lag_{(010)}$ and $\Lag_{(001)}$. As mentioned above, we are going to consider separately the ones with $n=0$ and $n=1$. We begin with $G_{3}\Lag_{(010)}=G_{3}\wedge\Phi^{a}\wedge\theta^{\star}_{~a}$. Its e.o.m. reads
\begin{equation}
\begin{split}
\delta(G_{3}\Lag_{(010)})=&\delta G_{3}\wedge\Phi^{a}\wedge\theta^{\star}_{~a}+G_{3}\wedge\delta\Phi^{a}\wedge\theta^{\star}_{~a} \\
=&\left(G_{3,\phi}\delta\phi-G_{3,X}\nabla_{z}\phi\nabla^{z}\delta\phi+\mathcal{O}\left(G_{3,[\Phi]}\right)\right)\wedge\Phi^{a}\wedge\theta^{\star}_{~a}+G_{3}\wedge \D\nabla^{a} \delta\phi\wedge\theta^{\star}_{~a} \\
=&\delta\phi\wedge\left(G_{3,\phi}\wedge\Phi^{a}\wedge\theta^{\star}_{~a}+\nabla^{z}\left(G_{3,X}\nabla_{z}\phi\wedge\Phi^{a}\wedge\theta^{\star}_{~a}\right)+\nabla^{a} \D\left(G_{3}\wedge\theta^{\star}_{~a}\right)\right)+\mathcal{O}\left(G_{3,[\Phi]}\right).
\end{split}
\label{eq:H1}
\end{equation}
Here, $\mathcal{O}\left(G_{3,[\Phi]}\right)$ encodes higher order terms coming from the dependence of $G_{3}$ in second order derivative scalars such as $[\Phi]$. In going from the second to the third line of (\ref{eq:H1}), we have used Stoke's theorem (\ref{eq:Stokes}) and assumed vanishing contributions at the boundary. We can expand the above expression further 
\begin{equation}
\begin{split}
\delta(G_{3}\Lag_{(010)})=&\delta\phi\wedge\left(G_{3,\phi}\wedge\Phi^{a}\wedge\theta^{\star}_{~a}+\nabla^{z}\left(G_{3,X}\nabla_{z}\phi\right)\wedge\Phi^{a}\wedge\theta^{\star}_{~a}+G_{3,X}\nabla_{z}\phi\wedge\dashuline{\nabla^{z}\Phi^{a}}\wedge\theta^{\star}_{~a}\right) \\
+&\delta\phi\wedge\left(\nabla^{a}\lp G_{3,\phi}\D\phi\rp\wedge\theta^{\star}_{~a}-\nabla^{a}\lp G_{3,X}\nabla_{z}\phi\rp\wedge\Phi^{z}\wedge\theta^{\star}_{~a}\right) \\
-&\delta\phi\wedge G_{3,X}\nabla_{z}\phi\wedge\dashuline{\nabla^{a}\Phi^{z}}\wedge\theta^{\star}_{~a}+\mathcal{O}\left(G_{3,[\Phi]}\right).
\end{split}
\label{eq:H1a}
\end{equation}
Now, using the commutation of covariant derivatives, (\ref{eq:id1}), we can eliminate the higher derivatives of the terms underlined with a dashed line, retaining only a curvature term. Moreover, from the previous argument for $p=0$, we must impose that $G_{3}=G_{3}(\phi,X)$ to avoid higher than two Euler-Lagrange equations. Consequently, any $\mathcal{O}\left(G_{3,[\Phi]}\right)$-term is zero. In the following computations, we will write directly the expanded expression after applying Stoke's theorem.

In conclusion, imposing $G_{3}=G_{3}(\phi,X)$ ensures that the e.o.m. from $\Lag_{(010)}$ remain 2nd order and, thus, automatically ghost-free. Therefore, we have found that the e.o.m. of
\begin{equation}
\Lag_{3}^{H}=G_{3}\wedge\Phi^{a}\wedge\theta^{\star}_{~a}=G_{3}(\phi,X)[\Phi]\eta
\label{eq:H3}
\end{equation}
are at most 2nd order. We have included the component expression of $\Lag_{3}^{H}$ to express that it can be directly linked with Horndeski's theory.

Subsequently, we study $E_{2}\Lag_{(001)}=E_{2}\wedge\Psi^{a}\wedge\theta^{\star}_{~a}$. Its e.o.m. reads
\begin{equation}
\begin{split}
\delta(E_{2}\Lag_{(001)})=&\delta E_{2}\wedge\Psi^{a}\wedge\theta^{\star}_{~a}+E_{2}\wedge\delta\Psi^{a}\wedge\theta^{\star}_{~a} \\
=&\delta\phi\wedge\left(E_{2,\phi}\wedge\Psi^{a}\wedge\theta^{\star}_{~a}+\nabla^{z}\left(E_{2,X}\nabla_{z}\phi\wedge\Psi^{a}\rp\wedge\theta^{\star}_{~a}\rp \\
-&\delta\phi\wedge\left(\nabla^{a}\left(E_{2}\wedge \D\phi\right)\wedge\theta^{\star}_{~a}+\D\left(E_{2}\nabla^{a}\phi\right)\wedge\theta^{\star}_{~a}\right)+\mathcal{O}\left(E_{2,[\Phi]}\right).
\end{split}
\label{eq:NH1}
\end{equation}
In this case, if we impose $E_{2}=E_{2}(\phi,X)$, then the e.o.m. directly remain 2nd order, obtaining
\begin{equation}
\Lag_{2}^{NH}=E_{2}\wedge\Psi^{a}\wedge\theta^{\star}_{~a}=-2XE_{2}(\phi,X)\eta.
\label{eq:NH2}
\end{equation}
Although in this case it is trivial to see that this Lagrangian belongs to $\Lag_{2}^{H}$, we will postpone this discussion until Sec. \ref{sec:Completeness}.

\item $p=2$

At this order, when $n=0$, we encounter two possible terms, $G_{4}\Lag_{(100)}=G_{4}\wedge\R^{ab}\wedge\theta^{\star}_{~ab}$ and $F_{4}\Lag_{(020)}=F_{4}\wedge\Phi^{a}\wedge\Phi^{b}\wedge\theta^{\star}_{~ab}$. A priori, the coefficients of each Lagrangian $G_{4}$ and $F_{4}$ are unrelated. We analyze each of them separately. Firstly, we have
\begin{equation}
\begin{split}
\delta(G_{4}\Lag_{(100)})=&\delta G_{4}\wedge\R^{ab}\wedge\theta^{\star}_{~ab} \\
=&\delta\phi\wedge\left(G_{4,\phi}\wedge\R^{ab}+\nabla^{z}\left(G_{4,X}\nabla_{z}\phi\right)\wedge\R^{ab}+G_{4,X}\nabla_{z}\phi\wedge\underline{\nabla^{z}\R^{ab}}\right)\wedge\theta^{\star}_{~ab}+\mathcal{O}\left(G_{4,[\Phi]}\right).
\end{split}
\label{eq:H2a}
\end{equation}

Secondly, we compute
\begin{equation}
\begin{split}
\delta(F_{4}\Lag_{(020)})=&\delta F_{4}\wedge\Phi^{a}\wedge\Phi^{b}\wedge\theta^{\star}_{~ab}+2F_{4}\wedge\delta\Phi^{a}\wedge\Phi^{b}\wedge\theta^{\star}_{~ab} \\
=&\delta\phi\wedge\left(F_{4,\phi}\wedge\Phi^{a}+\nabla^{z}\left(F_{4,X}\nabla_{z}\phi\right)\wedge\Phi^{a}+2F_{4,X}\nabla_{z}\phi\wedge\dashuline{\nabla^{z}\Phi^{a}}\right)\wedge\Phi^{b}\wedge\theta^{\star}_{~ab} \\
+&2\delta\phi\wedge\left(\nabla^{a}\left(F_{4,\phi}\D\phi\right)-\nabla^{a}\left(F_{4,X}\nabla_{z}\phi\right)\wedge\Phi^{z}\right)\wedge\Phi^{b}\wedge\theta^{\star}_{~ab} \\
+&2\delta\phi\wedge\left(-F_{4,X}\nabla_{z}\phi\wedge\dashuline{\nabla^{a}\Phi^{z}}\wedge\Phi^{b}\wedge\theta^{\star}_{~ab}+\D\left(F_{4}\right)\wedge\nabla^{a}\Phi^{b}\wedge\theta^{\star}_{~ab}\right) \\
+&2\delta\phi\wedge\left(\nabla^{a}\left(F_{4}\nabla_{z}\phi\right)\wedge\R^{bz}\wedge\theta^{\star}_{~ab}+F_{4}\nabla_{z}\phi\wedge\uline{\nabla^{a}\R^{bz}}\wedge\theta^{\star}_{~ab}\right)+\mathcal{O}\left(F_{4,[\Phi]}\right).
\end{split}
\label{eq:H2b}
\end{equation}
Subsequently, we can analyze the higher derivative terms, recalling that when all the indices are antisymmetrized there is no such a problematic term. As in the previous case, using the commutation of covariant derivatives, (\ref{eq:id1}), we can rewrite the dashed underlined terms of (\ref{eq:H2b}) as a curvature 2-form. Moreover, we can rearrange the solid underlined term of (\ref{eq:H2a}), recalling Bianchi identity, (\ref{eq:id2}), in order to compensate the corresponding one of (\ref{eq:H2b}). Doing so, we learn that these higher derivative terms only cancel each other if $F_{4}=G_{4,X}$. Additionally, 2nd order e.o.m. are only achieved if $G_{4}=G_{4}(\phi,X)$. Altogether, we have obtained that
\begin{equation}
\begin{split}
\Lag_{4}^{H}=& G_{4}\wedge\R^{ab}\wedge\theta^{\star}_{~ab}+G_{4,X}\wedge\Phi^{a}\wedge\Phi^{b}\wedge\theta^{\star}_{~ab} \\
=&\left(G_{4}R+G_{4,X}([\Phi]^2 -[\Phi^2])\right)\eta.
\end{split}
\label{eq:H4}
\end{equation}
is a second-order Lagrangian. Clearly, writing it in components, we recover the well-known $\Lag_{4}$ of Horndeski's theory. For the rest of the cases, the process of finding a Lagrangian with second order e.o.m. will be analogous to the one just described: using the commutation of covariant derivatives, (\ref{eq:id1}) one eliminates the higher derivatives of the dashed underlined terms, and, using Bianchi identity, (\ref{eq:id2}), and setting the appropriate coefficient, one cancels the unwanted parts underlined with solid lines. 

When $n=1$, we encounter $E_{3}\Lag_{(011)}=E_{3}\wedge\Phi^{a}\wedge\Psi^{b}\wedge\theta^{\star}_{~ab}$. Thus, the e.o.m. follows
\begin{equation}
\begin{split}
\delta(E_{3}\Lag_{(011)})=&\delta E_{3}\wedge\Phi^{a}\wedge\Psi^{b}\wedge\theta^{\star}_{~ab}+E_{3}\wedge\delta\Phi^{a}\wedge\Psi^{b}\wedge\theta^{\star}_{~ab}+E_{3}\wedge\Phi^{a}\wedge\delta\Psi^{b}\wedge\theta^{\star}_{~ab} \\
=&\delta\phi\wedge\left(E_{3,\phi}\wedge\Phi^{a}\wedge\Psi^{b}\wedge\theta^{\star}_{~ab}+\nabla^{z}\left(E_{3,X}\nabla_{z}\phi\right)\wedge\Phi^{a}\wedge\Psi^{b}\wedge\theta^{\star}_{~ab}\right) \\
+&\delta\phi\wedge\left(E_{3,X}\nabla_{z}\phi\wedge\left(\dashuline{\nabla^{z}\Phi^{a}}\wedge\Psi^{b}+\Phi^{a}\wedge\nabla^{\alpha}\Psi^{b}\right)+\nabla^{a}\left(D\phi\wedge\left(E_{3,\phi}\Psi^{b}-E_{3}\Phi^{b}\right)\right)\right)\wedge\theta^{\star}_{~ab} \\
+&\delta\phi\wedge\left(\nabla^{a}\left(E_{3,X}\nabla_{z}\phi\wedge\Psi^{b}\right)\wedge\Phi^{\alpha}\wedge\theta^{\star}_{~ab}-E_{3,X}\nabla_{z}\phi\wedge\dashuline{\nabla^{a}\Phi^{z}}\wedge\Psi^{b}\wedge\theta^{\star}_{~ab}\right) \\
-&\delta\phi\wedge\left(\nabla^{a}\left(E_{3}\wedge \D\phi\wedge\Phi^{b}\right)+\D\left(E_{3}\nabla^{a}\phi\wedge\Phi^{b}\right)\right)\wedge\theta^{\star}_{~ab}+\mathcal{O}\left(E_{3,[\Phi]}\right).
\end{split}
\label{eq:NH2a}
\end{equation}
Again, recalling the commutation of covariant derivatives, (\ref{eq:id1}), one can eliminate the higher derivatives of the terms underlined with a dash line. Consequently, if $E_{3}=E_{3}(\phi,X)$, the e.o.m. are at most 2nd order, reading
\begin{equation}
\Lag_{3}^{NH}=E_{3}\Phi^{a}\wedge\Psi^{b}\wedge\theta^{\star}_{~ab}=-E_{3}(\langle\Phi\rangle+2X[\Phi])\eta.
\label{eq:NH3}
\end{equation}
In Sec. \ref{sec:Completeness}, we will clarify the role of this Lagrangian.

\item $p=3$

Now, due to the fact that the calculations are going to be analogous to the ones presented in the previous cases, we show the results directly. Nevertheless, we include the complete computation in App. \ref{app:sEOM}. When $n=0$, there are two Lagrangians $\Lag_{(110)}$ and $\Lag_{(030)}$. We find that the following Lagrangian
\begin{equation}
\begin{split}
\Lag_{5}^{H}=& G_{5}\wedge\R^{ab}\wedge\Phi^{c}\wedge\theta^{\star}_{~abc}+\frac{1}{3}G_{5,X}\wedge\Phi^{a}\wedge\Phi^{b}\wedge\Phi^{c}\wedge\theta^{\star}_{~abc} \\
=&-2\left(G_{5}G^{ab}\Phi_{ab}-\frac{1}{6}G_{5,X}([\Phi]^3 -3[\Phi][\Phi^2]+2[\Phi^3])\right)\eta
\end{split}
\label{eq:H5}
\end{equation}
has second order e.o.m. Clearly, we have recovered $\Lag_{5}$ of Horndeski's theory.

When $n=1$, we can have $\Lag_{(101)}$ and $\Lag_{(021)}$. We obtain that
\begin{equation}
\begin{split}
\Lag_{4}^{NH}=& E_{4}\wedge\R^{ab}\wedge\Psi^{c}\wedge\theta^{\star}_{~abc}+E_{4,X}\wedge\Phi^{a}\wedge\Phi^{b}\wedge\Psi^{c}\wedge\theta^{\star}_{~abc} \\
=&-2\left(E_{4}\langle G\rangle-E_{4,X}\lp\langle\Phi^2\rangle-\langle\Phi\rangle[\Phi]+X([\Phi]^{2}-[\Phi^2]\rp\right)\eta.
\end{split}
\label{eq:NH4}
\end{equation}
has no Ostrogradski instabilities. Interestingly, this Lagrangian has structural similarities with $\Lag_{4}^{H}$ in (\ref{eq:H4}).

\item $p=4$

Finally, we analyze the case in which $p$ is maximum. Similarly to the previous case, we present directly the result and incorporate the details of the calculations in App. \ref{app:sEOM}. Considering first the Lagrangians with $n=1$, corresponding to $\Lag_{(111)}$ and $\Lag_{(031)}$, we find that they can be combined as
\begin{equation}
\begin{split}
\Lag_{5}^{NH}=& E_{5}\wedge\R^{ab}\wedge\Phi^{c}\wedge\Psi^{d}\wedge\theta^{\star}_{~abcd}+\frac{1}{3}E_{5,X}\wedge\Phi^{a}\wedge\Phi^{b}\wedge\Phi^{c}\wedge\Psi^{d}\wedge\theta^{\star}_{~abcd} \\
=&(E_{5}\left(4\left(\langle R_{ab}\Phi^{bc}\rangle+X[R\Phi]\right)-R\left(\langle\Phi\rangle+2X[\Phi]\right)+2\left(\langle R_{abcd}\Phi^{bd}\rangle-\langle R\rangle[\Phi]\right)\right) \\
+&\frac{1}{3}E_{5,X}((\langle\Phi^2\rangle[\Phi]-\langle\Phi^3\rangle)-3\langle\Phi\rangle([\Phi]^{2}-[\Phi^2])-2X([\Phi]^{3}-3[\Phi][\Phi^{2}]+2[\Phi^{3}]))\eta
\end{split}
\label{eq:NH5}
\end{equation}
to give second order e.o.m. One should notice that the four Lagrangians of Horndeski's theory have already appeared. However, until Sec. \ref{sec:Completeness}, we cannot conclude anything about this possible new Lagrangian.

When $n=0$, there are three possible terms: $\Lag_{(200)}$, $\Lag_{(120)}$ and $\Lag_{(040)}$. Nevertheless, the computation is equivalent to the previous cases in which we had two higher derivatives terms arising from each of the Lagrangians that cancel each other choosing the right coefficient. In this case, $\Lag_{(120)}$ will have two higher derivative terms that will be eliminated with two others coming from $\Lag_{(200)}$ and $\Lag_{(040)}$ (see details in App. \ref{app:sEOM}). At the end, we find that
\begin{equation}
\begin{split}
\Lag_{6}^{NH}=& E_{6}\wedge\R^{ab}\wedge\R^{cd}\wedge\theta^{\star}_{~abcd}+2E_{6,X}\R^{ab}\wedge\Phi^{c}\wedge\Phi^{d}\wedge\theta^{\star}_{~abcd}+\frac{1}{3}E_{6,XX}\Phi^{a}\wedge\Phi^{b}\wedge\Phi^{c}\wedge\Phi^{d}\wedge\theta^{\star}_{~abcd}\\
=&(E_{6}(R_{abcd}R^{abcd}-4R_{ef}R^{ef}+R^{2})+2E_{6,X}(R([\Phi]^{2}-[\Phi^2])-4R^{ab}([\Phi]\Phi_{ab}-\Phi^{2}_{ab})+2R^{abcd}\Phi_{ac}\Phi_{bd}) \\
+&\frac{1}{3}E_{6,XX}([\Phi]^4 -6[\Phi]^2[\Phi^2]+3[\Phi^2]^2+8[\Phi][\Phi^3]-6[\Phi^4]))\eta
\end{split}
\label{eq:NH6}
\end{equation}
has no higher derivatives in the e.o.m. In the following we will refer to the above theory as \emph{kinetic Gauss-Bonnet}. To the best of our knowledge, this general Lagrangian was \emph{not considered previously} in the literature. Nevertheless, the particular case when $E_{6}=E_{6}(\phi)$ describes a Gauss-Bonnet (GB) gravity coupled with a scalar function, which is a model extensively studied in the literature, see e.g. \cite{Nojiri:2005vv}. In fact, in Ref. \cite{Kobayashi:2011nu}, it was claimed that $f(\phi)GB$ is already contained in Horndeski's theory. In the next section, we will investigate if this result holds for the more general Lagrangian presented in (\ref{eq:NH6}).
\end{enumerate}

In order to conclude the analysis of the scalar e.o.m., we are going to study a general $\Lag_{(lmn)}$ in $D$-dimensions. This calculation will help us understand how the higher than second order terms in the e.o.m. can be cancelled in $D$-dimensions and what is the origin of the concrete numerical factors obtained. In particular, we are looking for relations between different coefficients in front of each $\Lag_{(lmn)}$ that build up \emph{automatically ghost-free} combinations of Lagrangians, as we have done before. In this computation, since we are considering a general case, we will use the general notation for the  coefficient $\alpha_{lmn}$. As we have discussed for the e.o.m. in $4D$, a dependence of $\alpha_{lmn}$ in second derivatives or higher induces higher than two derivatives terms in the e.o.m. that cannot be cancelled since they are proportional to the original Lagrangian, e.g. $\nabla^{z}\nabla_{z}(\alpha_{lmn,[\Phi]})\Lag_{(lmn)}$. For that reason, we impose from the beginning that $\alpha_{lmn}=\alpha_{lmn}(\phi,X)$. Also, we will be mostly interested in the remaining higher derivative terms. Therefore, we will try to keep the rest as simple as possible. Making a variation with respect to the scalar field, we obtain
\begin{equation}
\begin{split}
\delta(\alpha_{lmn}\Lag_{(lmn)})=&\delta \alpha_{lmn}\wedge\Lag_{(lmn)}+\alpha_{lmn}\wedge\delta\Lag_{(lmn)} \\
=&\delta \alpha_{lmn}\wedge\Lag_{(lmn)}+m\alpha_{lmn}\wedge\delta\Phi^{a}\wedge[\Lag_{(l(m-1)n)}]_{a}+n\alpha_{lmn}\wedge\delta\Psi^{a}\wedge[\Lag_{(lm(n-1))}]_{a} \\
=&\delta\phi\wedge\Big(\big(\alpha_{lmn,\phi}+\nabla^{z}(\alpha_{lmn,X}\nabla_{z}\phi)\big)\wedge\Lag_{(lmn)}+\alpha_{lmn,X}\nabla_{z}\phi\big(l\uline{\nabla^{z}\R^{ab}}\wedge[\Lag_{((l-1)mn)}]_{ab} \\
+&m\uline{\nabla^{z}\Phi^{a}}\wedge[\Lag_{(l(m-1)n)}]_{a}+n\nabla^{z}\Psi^{a}\wedge[\Lag_{(lm(n-1))}]_{a}\big)+m\big(\nabla^{a}(\alpha_{lmn,\phi}\D\phi\wedge[\Lag_{(l(m-1)n)}]_{a}) \\
-&\alpha_{lmn,X}\nabla_{z}\phi\uline{\nabla^{a}\Phi^{z}}\wedge[\Lag_{(l(m-1)n)}]_{a}-\Phi^{z}\wedge\nabla^{a}(\alpha_{lmn,X}\nabla_{z}\phi[\Lag_{(l(m-1)n)}]_{a})\big) \\
+&m(m-1)\big(\alpha_{lmn}\nabla_{z}\phi \uline{\nabla^{a}\R^{bz}}\wedge[\Lag_{(l(m-2)n)}]_{ab}+\R^{bz}\wedge\nabla^{a}(\alpha_{lmn}\nabla_{z}\phi[\Lag_{(l(m-2)n)}]_{ab})\big) \\
+&mn\nabla^{a}(\alpha_{lmn}\D\Psi^{b}\wedge[\Lag_{(l(m-1)(n-1))}]_{ab})+n\alpha_{lmn}\wedge\delta\Psi^{a}\wedge[\Lag_{(lm(n-1))}]_{a}\Big).
\end{split}
\end{equation}
where $[\Lag_{(lmn)}]_{a_{1}\cdots a_{k}}$ indicates that the first $k$ indices of the Hodge dual basis of $\Lag_{(lmn)}$ are free.
From the above result, we can see that there are four higher derivative terms, underlined with a solid line, 
\begin{align}
&m\alpha_{lmn,X}\nabla_{z}\phi\nabla^{z}\Phi^{a}\wedge[\Lag_{(l(m-1)n)}]_{a}, \\
-&m\alpha_{lmn,X}\nabla_{z}\phi\nabla^{a}\Phi^{z}\wedge[\Lag_{(l(m-1)n)}]_{a}, \\
-&2l\alpha_{lmn,X}\nabla_{z}\phi\nabla^{a}\R^{bz}\wedge[\Lag_{((l-1)mn)}]_{ab}, \\
&m(m-1)\alpha_{lmn}\nabla_{z}\phi\nabla^{a}\R^{bz}\wedge[\Lag_{(l(m-2)n)}]_{ab},
\end{align}
where we have already used Bianchi identity (\ref{eq:id2}) to rewrite the third one. Clearly, the first two terms safely add up to give a Riemann tensor, since they form a commutator of covariant derivatives, cf. (\ref{eq:id1}). However, the last two terms must be canceled with extra Lagrangians with different $(lmn)$. These new terms added will contribute also with other higher derivative terms. We can repeat this process iteratively until we reach a Lagrangian that does not contribute with extra higher order terms. Thus, the resulting combination that avoids higher order derivatives is
\begin{equation}
\Lag^{2^{nd}}(\alpha_{lmn})=\alpha_{lmn}\Lag_{(lmn)}+\sum_{j=1}^{l}\alpha_{(l-j)(m+2j)n}\Lag_{((l-j)(m+2j)n)}+\sum_{k=1}^{m/2}\alpha_{(l+k)(m-2k)n}\Lag_{((l+k)(m-2k)n)},
\label{eq:lagD}
\end{equation}
where the coefficients are related iteratively by
\begin{align}
\alpha_{(l-j)(m+2j)n}&=\frac{2(l-(j+1))}{(m+2j)(m+2j-1)}\frac{\partial(\alpha_{(l-(j-1))(m+2(j-1))n})}{\partial X}, \\
\alpha_{(l+k)(m-2k)n}&=\frac{(m-2(k-1))(m-1-2(k-1))}{2(l+k)}\int\alpha_{(l+(k-1))(m-2(k-1))n}dX. 
\end{align}
With these general expression, we can easily derive, for instance, Horndeski Lagrangians, i.e. (\ref{eq:H2}), (\ref{eq:H3}), (\ref{eq:H4}) and (\ref{eq:H5}). Although in $D$-dimensions there are $(D+1)(D+2)/2$ possible Lagrangians $\Lag_{(lmn)}$, the above result indicates that there are only $2D+1$ independents linear combinations giving second order e.o.m. 

\subsection{Vielbein Equations of Motion}
\label{subsec:vEOM}

Subsequently, we proceed to compute the corresponding equations of motion for the frame field $\theta^{a}$. In doing so, we will apply a second order approach, which means that we define the 1-form connection as a function of the vielbein only, i.e. $\omega=\omega(\theta)$. This is equivalent to the metric formalism, in which it is assumed that the only dynamical degrees of freedom are contained in the metric, i.e. $\Gamma(g)$. Alternatively, one could have chosen a first order approach (or Palatini formalism), in which the connection 1-form and the vielbein are independent variables. As it will become clear in the computation, our method can be easily extrapolated to that situation. Having developed the general framework to taking variations with respect to differential forms, we will directly study the general $D$-dimensional case.

A first point to consider is how to relate the 1-form connection and the vielbein. We can do this by fixing a metric-compatible and torsionless connection, i.e. $\omega_{ab}=-\omega_{ba}$ and $T^{a}=\D\theta^{a}=0$ respectively. If we have two different connections $\omega^{ab}$ and $\tilde{\omega}^{ab}$, arising from $\theta^{a}$ and $\tilde{\theta}^{a}$, we find that they are related by
\begin{equation}
\tilde{\omega}^{ab}=\omega^{ab}+\frac{1}{2}\lp\mathfrak{i}_{\tilde{e}^{b}}(\D\tilde{\theta}^{a})-\mathfrak{i}_{\tilde{e}^{a}}(\D\tilde{\theta}^{b})+\mathfrak{i}_{\tilde{e}^{a}}(\mathfrak{i}_{\tilde{e}^{b}}(\D\tilde{\theta}_{c}))\tilde{\theta^{c}}\rp,
\end{equation}
which is nothing but the differential form version of the usual torsionless spin connection in supergravity \cite{Ortin:2004ms}. Alternatively, it is the non-coordinate analog of the relation between two different Levi-Civita Connections \cite{Wald:1984rg}. To clarify this formal definition, we include a more detailed discussion in App. \ref{app:Notation}. Therefore, if we define a perturbed connection $\delta\omega^{ab}$, to linear order in the perturbations of $\delta\theta^{a}$, it will be
\begin{equation}
\delta\omega^{ab}=\frac{1}{2}\lp\mathfrak{i}_{e^{b}}\lp\D\delta\theta^{a}\rp-\mathfrak{i}_{e^{a}}\lp\D\delta\theta^{b}\rp+\mathfrak{i}_{e^{a}}\lp\mathfrak{i}_{e^{b}}\lp\D\delta\theta_{c}\rp\rp\theta^{c}\rp=\nabla^{b}\delta\theta^{a}-\nabla^{a}\delta\theta^{b}.
\label{eq:PertConnection}
\end{equation}
Thus, we have a relation that links $\delta\omega$ with $\delta\theta$.

As a consequence, the vielbein e.o.m. are going to be given by
\begin{equation}
\begin{split}
\delta \Lag&=\delta\theta^{a}\wedge\frac{\delta\Lag}{\delta\theta^{a}}+\delta\theta^{a}\wedge\frac{\delta \omega^{bc}}{\delta\theta^{a}}\wedge\frac{\delta\Lag}{\delta\omega^{bc}} \\
&=\delta\theta^{a}\wedge\frac{\delta\Lag}{\delta\theta^{a}}-\delta\theta^{a}\wedge\nabla^{b}\left(\frac{\delta\Lag}{\delta\omega^{ab}}\right)+\delta\theta^{b}\wedge\nabla^{a}\left(\frac{\delta\Lag}{\delta\omega^{ab}}\right),
\end{split}
\label{eq:veom}
\end{equation}
where in the second line we have used the variation of the 1-form connection given in (\ref{eq:PertConnection}) and integrated by parts, neglecting the surface terms. Nevertheless, if we want to analyze the higher derivatives only, we do not need to calculate all the terms. First, we should notice that $\theta^{a}$ appears linearly (with an exterior product) in $\Phi^{a}$, $\Psi^{a}$ and $\theta^{\star}_{a_{1}\cdots a_{k}}$. Consequently, since our Lagrangian does not contain higher than second derivatives by construction, the variation with respect to $\theta^{a}$ is not going to introduce them in the dynamical equations. Next, we should notice that the 1-form connection appears linearly in the second derivative of the scalar field, i.e. $\Phi^{a}=d\nabla^{a}\phi+\omega^{ab}\nabla_{b}\phi$, and through a covariant exterior derivative in the 2-form curvature, i.e. $\R^{ab}=\D\omega^{ab}$. For these reasons, the variation with respect to the connection, $\delta_{\omega}$, takes the form
\begin{equation}
\begin{split}
\delta_{\omega}(\alpha_{lmn}\Lag_{(lmn)})&=l\delta_{\omega}(\R^{ab})\wedge \alpha_{lmn}[\Lag_{((l-1)mn)}]_{ab}+m\delta_{\omega}(\Phi^{a})\wedge \alpha_{lmn}[\Lag_{(l(m-1)n)}]_{a} \\
&=\delta\omega^{ab}\wedge(l(\alpha_{lmn,\phi}\D\phi\wedge[\Lag_{((l-1)mn)}]_{ab}-\alpha_{lmn,X}\nabla_{z}\phi\Phi^{z}\wedge[\Lag_{((l-1)mn)}]_{ab} \\
&+m\alpha_{lmn}\R^{cd}\nabla_{d}\phi\wedge[\Lag_{((l-1)(m-1)n)}]_{abc}+n\alpha_{lmn}\Phi^{c}\wedge\D\phi\wedge[\Lag_{((l-1)m(n-1))}]_{abc}) \\
&+m\alpha_{lmn}\nabla_{b}\phi\wedge[\Lag_{(l(m-1)n)}]_{a}),
\end{split}
\end{equation}
where we are taking $\alpha_{lmn}=\alpha_{lmn}(\phi,X)$ for the same arguments discussed previously (see Sec. \ref{subsec:sEOM}). Then, to obtain the contribution to the e.o.m., we only need to apply the covariant derivative. The first term will be
\begin{equation}
\begin{split}
\delta\theta^{a}\wedge\nabla^{b}\bigg(\frac{\delta(\alpha_{lmn}\Lag_{(lmn)})}{\delta\omega^{ab}}&\bigg)=
\delta\theta^{a}\wedge\Big(l\big(\nabla^{b}(\alpha_{lmn,\phi}\D\phi\wedge[\Lag_{((l-1)mn)}]_{ab})-\Phi^{z}\wedge\nabla^{b}(\alpha_{lmn,X}\nabla_{z}\phi[\Lag_{((l-1)mn)}]_{ab}) \\
&-\alpha_{lmn,X}\nabla_{z}\phi\uline{\nabla^{b}\Phi^{z}}\wedge[\Lag_{((l-1)mn)}]_{ab}+m\R^{cz}\wedge \nabla^{b}(\alpha_{lmn}\nabla_{z}\phi[\Lag_{((l-1)(m-1)n)}]_{abc}) \\
&+m\alpha_{lmn}\nabla_{z}\phi\uline{\nabla^{b}\R^{cz}}\wedge[\Lag_{((l-1)(m-1)n)}]_{abc}+\nabla^{b}(n\alpha_{lmn}\Phi^{c}\wedge\D\phi\wedge[\Lag_{((l-1)m(n-1))}]_{abc})\big) \\
&+m\big(\nabla^{z}(\alpha_{lmn}\nabla_{z}\phi)\wedge[\Lag_{(l(m-1)n)}]_{a}+n\alpha_{lmn}\nabla_{z}\phi\nabla^{z}\Psi^{b}\wedge[\Lag_{(l(m-1)(n-1))}]_{ab} \\
&+l\alpha_{lmn}\nabla_{z}\phi\uline{\nabla^{z}\R^{bc}}\wedge[\Lag_{((l-1)(m-1)n)}]_{abc}+(m-1)\alpha_{lmn}\nabla_{z}\phi\uline{\nabla^{z}\Phi^{b}}\wedge[\Lag_{(l(m-2)n)}]_{ab}\big)\Big)
\end{split}
\label{eq:termI}
\end{equation}
and the second one will be
\begin{equation}
\begin{split}
\delta\theta^{b}\wedge\nabla^{a}\bigg(\frac{\delta(\alpha_{lmn}\Lag_{(lmn)})}{\delta\omega^{ab}}&\bigg)=
\delta\theta^{b}\wedge\Big(l\big(\nabla^{a}(\alpha_{lmn,\phi}\D\phi\wedge[\Lag_{((l-1)mn)}]_{ab})-\Phi^{z}\wedge\nabla^{a}(\alpha_{lmn,X}\nabla_{z}\phi[\Lag_{((l-1)mn)}]_{ab}) \\
&\hspace{-5pt}-\alpha_{lmn,X}\nabla_{z}\phi\uline{\nabla^{a}\Phi^{z}}\wedge[\Lag_{((l-1)mn)}]_{ab}+m\R^{cz}\wedge \nabla^{a}(\alpha_{lmn}\nabla_{z}\phi[\Lag_{((l-1)(m-1)n)}]_{abc}) \\
&\hspace{-5pt}+m\alpha_{lmn}\nabla_{z}\phi\uline{\nabla^{a}\R^{cz}}\wedge[\Lag_{((l-1)(m-1)n)}]_{abc}+\nabla^{a}(n\alpha_{lmn}\Phi^{c}\wedge\D\phi\wedge[\Lag_{((l-1)m(n-1))}]_{abc})\big) \\
&\hspace{-5pt}+m\nabla^{a}\big(\alpha_{lmn}\nabla_{b}\phi\wedge[\Lag_{(l(m-1)n)}]_{a}\big)\Big),
\end{split}
\label{eq:termII}
\end{equation}
where we have used the fact that a covariant derivative contracted with $\theta^{\star}_{a_{1}\cdots a_{k}}$, in this case shown by the indices $[\Lag]_{a_{1}\cdots a_{k}}$, acting on another term contracted with $\theta^{\star}_{a_{1}\cdots a_{k}}$ too is not generating higher derivatives because the indices are antisymmetric. We observe that (\ref{eq:termI}) has the following higher derivative terms, underlined with a solid line,
\begin{align}
&-l\alpha_{lmn,X}\nabla_{z}\phi\nabla^{b}\Phi^{z}\wedge[\Lag_{((l-1)mn)}]_{ab} \label{eq:V1}\\
&+lm\alpha_{lmn}\nabla_{z}\phi\nabla^{b}\R^{cz}\wedge[\Lag_{((l-1)(m-1)n)}]_{abc} \label{eq:V2}\\
&+ml\alpha_{lmn}\nabla_{z}\phi\nabla^{z}\R^{bc}\wedge[\Lag_{((l-1)(m-1)n)}]_{abc} \label{eq:V3}\\
&+m(m-1)\alpha_{lmn}\nabla_{z}\phi\nabla^{z}\Phi^{b}\wedge[\Lag_{(l(m-2)n)}]_{ab} \label{eq:V4}
\end{align}
On the other hand, (\ref{eq:termII}) has the following higher derivative terms
\begin{align}
&-l\alpha_{lmn,X}\nabla_{z}\phi\nabla^{b}\Phi^{z}\wedge[\Lag_{((l-1)mn)}]_{ab} \label{eq:V5}\\
&+lm\alpha_{lmn}\nabla_{z}\phi\nabla^{b}\R^{cz}\wedge[\Lag_{((l-1)(m-1)n)}]_{abc} \label{eq:V6}
\end{align}
where we have exchanged $a$ and $b$ in order to get more similar expressions to the above ones and introduced the relative sign between (\ref{eq:termI}) and (\ref{eq:termII}) appearing in (\ref{eq:veom}). Using the usual commutation of indices it is straightforward to see that the sum of (\ref{eq:V2}), (\ref{eq:V3}), and  (\ref{eq:V6}) is zero. Thus, we are led with only two higher derivatives terms
\begin{align}
&-2l\alpha_{lmn,X}\nabla_{z}\phi\nabla^{b}\Phi^{z}\wedge[\Lag_{((l-1)mn)}]_{ab} \\
&+m(m-1)\alpha_{lmn}\nabla_{z}\phi\nabla^{z}\Phi^{b}\wedge[\Lag_{(l(m-2)n)}]_{ab} 
\end{align}
where the first one correspond to the sum of (\ref{eq:V1}) and (\ref{eq:V5}), and the second one is (\ref{eq:V4}). 

As a consequence, in order to eliminate the higher derivative terms of the e.o.m. we have to add counter terms iteratively, equivalently to the previous case of scalar e.o.m. In fact, we find that the result is the same, i.e. the Lagrangian given by (\ref{eq:lagD}). It is interesting that in the scalar case the higher terms come from the derivatives of the curvature while in the vielbein case they appear from third derivatives of the field. This result has important consequences because it means that the Lagrangian (\ref{eq:lagD}) is a scalar-tensor theory in $D$-dimensions whose Euler-Lagrange equation is second order in derivatives.

\section{Completeness of the Formulation}
\label{sec:Completeness}

After computing the e.o.m., we must investigate what is the role of those Lagrangians $\Lag_{i}^{NH}$ that cannot be \emph{directly} linked with Horndeski's theory. To accomplish this task, we must first study if there are redundancies in our basis of Lagrangians, meaning that different $\Lag_{(lmn)}$ give the same e.o.m. This could happen if two Lagrangians are related by an exact form, the differential form analog of a total derivative, or by an algebraic identity. Consequently, we will study first the space of exact forms defined by $\Lag_{(lmn)}$. Then, we will analyze algebraic identities due to the antisymmetry of the Hodge dual basis. In both cases, we will start the analysis in $D$-dimensions and then particularize for $4D$. Finally, we will apply all these identities, which will act as constraints linking different Lagrangians, to conclude what is the most general basis and what are their corresponding combinations with second order e.o.m.

\subsection{Exact Forms}
\label{subsec:ExactForms}

We begin the analysis of the completeness of the formulation by computing the possible exact forms. An exact form is a $q$-form defined as an exterior derivative of a $(q-1)$-form, i.e. $\omega_{q}=d\omega_{q-1}$. It is important to consider them because by applying Stoke's theorem (\ref{eq:Stokes}), assuming no contribution at the boundary, they do not contribute to the e.o.m. We can build the space of exact $D$-forms in an analogous way to $\Lag_{(lmn)}$ in (\ref{eq:L}). However, since the exterior derivative is a mapping from $q$-forms to $(q+1)$-forms, we should start with an $\Lag_{(lmn)}$ satisfying $p\leq D-1$. Moreover, we should be aware that the final outcome must be part of the basis of Lagrangians in order to have a \emph{closed} set. For that reason, we cannot directly consider the reduction of $\Lag_{(lmn)}$ with an interior product, which is a mapping from $q$-forms to $(q-1)$-forms, because terms such as $\D(\mathfrak{i}_{\nabla\phi}\R^{ab})$ or $\D(\mathfrak{i}_{\nabla\phi}\Phi^{a})$ do not belong to the set of $\Lag_{(lmn)}$, $\Lag_{(\bar{l}m0)}$ and $\Lag_{(l\bar{m}0)}$ presented in (\ref{eq:L}), (\ref{eq:Lbarl}) and (\ref{eq:Lbarm}) respectively. Thus, $\D(\mathfrak{i}_{\nabla\phi}\mathcal{L}_{(lmn)})$ cannot be used.

Alternatively, we could define the space of exact forms by contracting one of the indices of the Hodge dual basis with a gradient of the scalar field. This is because the Hodge dual basis $\theta^{\star}_{a_{1}\cdots a_{k}}$ is a $(D-k)$-form. Thus, adding one index is equivalent to reducing one order in the differential form, which is exactly what we were looking for. This could be seen too as applying the interior product $\mathfrak{i}_{\nabla\phi}$ only to $\theta^{\star}_{a_{1}\cdots a_{k}}$. Noticeably, applying the interior product to $\Psi^{a}$, one obtains the same result with an extra $-2X$ factor. For all these arguments, we find that the appropriate space of exact forms is
\begin{equation}
\D\Lag^{D-1}_{(lmn)}[G_{i}]=\D\left(G_{i}\bigwedge_{i=1}^{l}\R^{a_{i}b_{i}}\wedge\bigwedge_{j=1}^{m}\Phi^{c_{j}}\wedge\bigwedge_{k=1}^{n}\Psi^{d_{k}}\wedge\left.\theta\right.^{\star}_{~ea_{1}b_{1}\cdots a_{l}b_{l}c_{1}\cdots c_{m}d_{1}\cdots d_{n}}\nabla^{e}\phi\right),
\label{eq:ExactForm}
\end{equation}
where the contraction of the last index of the Hodge dual ensures that we have a (D-1)-form inside the exterior derivative. One should notice that the above expression identically vanishes if $n\neq0$. This is because $\Psi^{a}=\nabla^{a}\phi D\phi$ and the antisymmetry of the Hodge dual basis kills it, i.e. $\theta^{\star}_{~ab}\nabla^{a}\phi\nabla^{b}\phi=0$. This cancellation will happen again for any Lagrangian containing $\Psi$. In contrast to the previous expression, this exact form only generates terms belonging to $\Lag_{(lmn)}$, $\Lag_{(\bar{l}mn)}$ and $\Lag_{(l\bar{m}n)}$.

Since we have already worked out the action of $\D$ in all of the building blocks for the e.o.m., we can easily expand the exact form (\ref{eq:ExactForm})
 \begin{equation}
 \begin{split}
\D\Lag_{(lmn)}^{D-1}[G_{i}]=G_{i,\phi}\Lag_{(lm(n+1))}-G_{i,X}\Lag_{(l(\overline{m+1})n)}+G_{i}\left(\Lag_{(l(m+1)n)}-m\Lag_{((\overline{l+1})(m-1)n)}-n\Lag_{(l(m+1)n)}\right).
\end{split}
\label{eq:Dexactform}
\end{equation}
This expression sets the general shape of an exact form. It implies that there is a linear dependence between some $\Lag_{(lmn)}$ and their contracted version $\Lag_{(\bar{l}mn)}$ and $\Lag_{(l\bar{m}n)}$. In $D=4$, there are six non-zero exact forms, i.e.
\begin{align}
\D\mathcal{L}_{(000)}^{D-1}[G_{2}]=&G_{2,\phi}\Lag_{(001)}-G_{2,X}\Lag_{(0\bar{1}0)}+G_{2}\Lag_{(010)}, \label{eq:D1}\\
\D\mathcal{L}_{(010)}^{D-1}[G_{3}]=&G_{3,\phi}\Lag_{(011)}-G_{3,X}\Lag_{(0\bar{2}0)}+G_{3}\left(\Lag_{(020)}-\Lag_{(\bar{1}00)}\right), \label{eq:D2}\\
\D\mathcal{L}_{(100)}^{D-1}[G_{4}]=&G_{4,\phi}\Lag_{(101)}-G_{4,X}\Lag_{(1\bar{1}0)}+G_{4}\Lag_{(110)}, \label{eq:D3}\\
\D\mathcal{L}_{(020)}^{D-1}[F_{4}]=&F_{4,\phi}\Lag_{(021)}-F_{4,X}\Lag_{(0\bar{3}0)}+F_{4}\left(\Lag_{(030)}-2\Lag_{(\bar{1}10)}\right), \label{eq:D4}\\
\D\mathcal{L}_{(110)}^{D-1}[G_{5}]=&G_{5,\phi}\Lag_{(111)}-G_{5,X}\Lag_{(1\bar{2}0)}+G_{5}\left(\Lag_{(120)}-\Lag_{(\bar{2}00)}\right), \label{eq:D5}\\
\D\mathcal{L}_{(030)}^{D-1}[F_{5}]=&F_{5,\phi}\Lag_{(031)}-F_{5,X}\Lag_{(0\bar{4}0)}+F_{5}\left(\Lag_{(040)}-3\Lag_{(\bar{1}20)}\right). \label{eq:D6}
\end{align}
In App. \ref{app:ExactForms}, we include the explicit computation of each of them. We will make full use of these expressions in the next subsections.
\subsection{Antisymmetric Degeneracies}
\label{subsec:AntisymDegeneracies}

To continue the analysis of possible degeneracies in the set of Lagrangians, we consider now identities derived from the antisymmetry of the Hodge dual basis $\theta^{\star}_{a_{1}\cdots a_{n}}$. Using its definition (\ref{eq:HD}), it is easy to prove that Hodge dual bases with a different number of indices are related by 
 \begin{equation}
\theta^{a}\wedge\theta^{\star}_{b_{1}\cdots b_{k}}=\delta^{a}_{b_{k}}\theta^{\star}_{b_{1}\cdots b_{k-1}}-\delta^{a}_{b_{k-1}}\theta^{\star}_{b_{1}\cdots b_{k-2}b_{k}}+\cdots+(-1)^{k-1}\delta^{a}_{b_{1}}\theta^{\star}_{b_{2}\cdots b_{k-1}}.
\end{equation}
In fact, this identity, rewritten in components via the totally antisymmetric tensor, was used by Horndeski to rewrite its equations of motion in \cite{Horndeski:1974wa}. Also, it was utilized by Ref. \cite{Charmousis:2008kc} in their study of Lovelock's theories. 

\begin{figure}[t!]
\begin{center}
\includegraphics{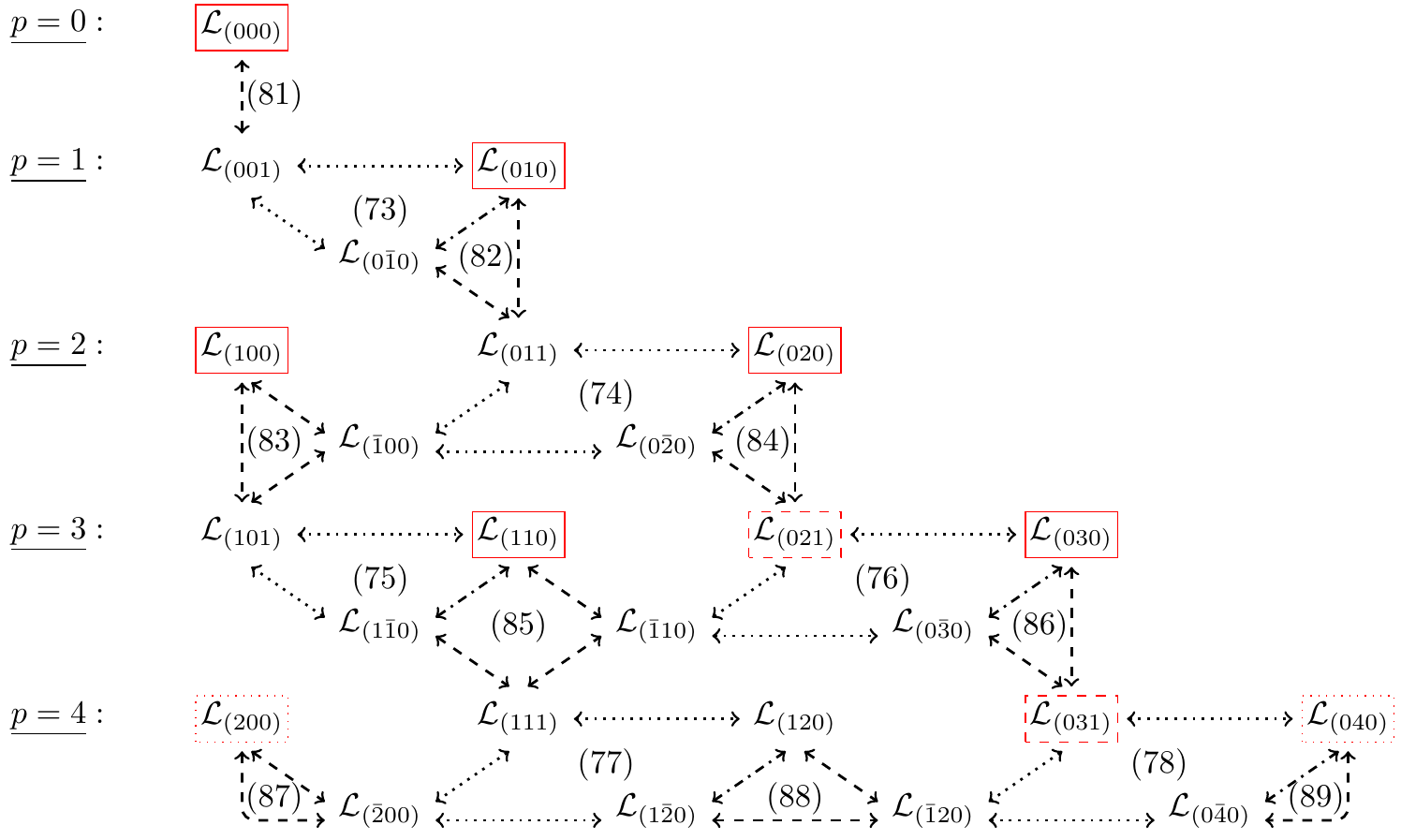}
\end{center}
\vspace{-10pt}
\caption{Summary of the interconnections between different Lagrangians $\Lag_{(lmn)}$, $\Lag_{(\bar{l}mn)}$ and $\Lag_{(l\bar{m}n)}$, defined respectively by (\ref{eq:L}), (\ref{eq:Lbarl}) and (\ref{eq:Lbarm}) in $4D$. A close set of arrows indicates that the Lagrangians in the vertices are related by the identity referred in the interior, which can be either an exact form (\ref{eq:D1}-\ref{eq:D6}), presented with dotted arrows, or an algebraic antisymmetric identity (\ref{eq:L1}-\ref{eq:L9}), plotted with dashed arrows. Here, a dash-dotted arrow indicates that two Lagrangians are related by both types of identities. In total, there are 10 independent Lagrangians. In the figure, we show a possible choice, framing each term in a rectangle, corresponding to Horndeski theory (red rectangles), Beyond Horndeski's $G^3$ (red dashed rectangles) and $\Lag_{(200)}$ and $\Lag_{(040)}$ (red dotted rectangles). Finally, we emphasize the structure by levels indicating in the left the number of building blocks $p\equiv2l+m+n$.}
\label{fig:Interconnection}
\end{figure}

Applying this identity, we observe that it can be used to relate different Lagrangians. We conclude that a general $\Lag_{(lmn)}$ with $n=1$ can be related with other Lagrangians with $n=0$ through
 \begin{equation}
\Lag_{(lm1)}=-2l\Lag_{(\bar{l}m0)}-m\Lag_{(l\bar{m}0)}-2X\Lag_{(lm0)}.
\label{eq:Ln}
\end{equation}
Since we are working in $D=4$, we can find nine new identities, i.e.
\begin{align}
\Lag_{(001)}=&-2X\Lag_{(000)}, \label{eq:L1} \\
\Lag_{(011)}=&-\Lag_{(0\bar{1}0)}-2X\Lag_{(010)}, \label{eq:L2} \\
\Lag_{(101)}=&-2\Lag_{(\bar{1}00)}-2X\Lag_{(100)}, \label{eq:L3} \\
\Lag_{(021)}=&-2\Lag_{(0\bar{2}0)}-2X\Lag_{(020)}, \label{eq:L4} \\
\Lag_{(111)}=&-2\Lag_{(\bar{1}10)}-\Lag_{(1\bar{1}0)}-2X\Lag_{(110)}, \label{eq:L5} \\
\Lag_{(031)}=&-3\Lag_{(0\bar{3}0)}-2X\Lag_{(030)}. \label{eq:L6} \\
\Lag_{(201)}=&-4\Lag_{(\bar{2}00)}-2X\Lag_{(200)}=0, \label{eq:L7} \\
\Lag_{(121)}=&-2\Lag_{(\bar{1}20)}-2\Lag_{(1\bar{2}0)}-2X\Lag_{(120)}=0, \label{eq:L8} \\
\Lag_{(041)}=&-4\Lag_{(0\bar{4}0)}-2X\Lag_{(040)}=0. \label{eq:L9} 
\end{align}
These 9 new relations together with the previous 6 exact forms add up to a total of 15 constraints. Thus, if we sum all the possible $\Lag_{(lmn)}$ in $D=4$, i.e. 15, and all the possible $\Lag_{(\bar{l}mn)}$ and $\Lag_{(l\bar{m}n)}$, i.e. 10, we are left with 10 independent Lagrangians. Among them, there is a certain freedom in the choice, that we summarize in Fig. \ref{fig:Interconnection}. From all the possibilities, one could choose to have the six terms of Horndeski theory, i.e. $\Lag_{(000)}$, $\Lag_{(010)}$, $\Lag_{(100)}$, $\Lag_{(020)}$, $\Lag_{(110)}$ and $\Lag_{(030)}$; the two of Beyond Horndeski's $G^3$, i.e. $\Lag_{(021)}$ and $\Lag_{(031)}$; and two additional terms, which we choose to be $\Lag_{(040)}$ and the Gauss-Bonnet term $\Lag_{(200)}$. In the next subsection, we will see how these constraints affect the Lagrangians that we have found with second order e.o.m., which are the relevant ones at the end.
\subsection{Relations between Second-Order Theories}
\label{subsec:BeyondHorndeski}

After computing all the relations that connect different terms in our basis of Lagrangians, the key question is: do these Lagrangians represent any viable/second-order sector different from Horndeski? To analyze this point, we will look at the combinations of Lagrangians whose e.o.m. are second order. We recall that in Sec. \ref{sec:EOM} we have found two sets of Lagrangians $\Lag^{H}_{i}$ and $\Lag_{i}^{NH}$ satisfying the latter condition. We summarize the first set of Lagrangians in
\begin{align}
\Lag_{2}^{H}[G_{2}]=&G_{2}\Lag_{(000)}, \label{eq:LH2} \\
\Lag_{3}^{H}[G_{3}]=&G_{3}\Lag_{(010)}, \label{eq:LH3} \\
\Lag_{4}^{H}[G_{4}]=&G_{4}\Lag_{(100)}+G_{4,X}\Lag_{(020)}, \label{eq:LH4} \\
\Lag_{5}^{H}[G_{5}]=&G_{5}\Lag_{(110)}+\frac{1}{3}G_{5,X}\Lag_{(030)}, \label{eq:LH5} 
\intertext{which is nothing but the differential form version of Horndeski's theory. We can englobe the whole set in $\Lag^{H}=\sum_{i=2}^{5}\Lag_{i}^{H}$. The second set we found was}
\Lag_{2}^{NH}[E_{2}]=&E_{2}\Lag_{(001)}, \label{eq:LNH2} \\
\Lag_{3}^{NH}[E_{3}]=&E_{3}\Lag_{(011)}, \label{eq:LNH3} \\
\Lag_{4}^{NH}[E_{4}]=&E_{4}\Lag_{(101)}+E_{4,X}\Lag_{(021)}, \label{eq:LNH4} \\
\Lag_{5}^{NH}[E_{5}]=&E_{5}\Lag_{(111)}+\frac{1}{3}E_{5,X}\Lag_{(031)}, \label{eq:LNH5} \\
\Lag_{6}^{NH}[E_{6}]=&E_{6}\Lag_{(200)}+2E_{6,X}\Lag_{(120)}+\frac{1}{3}E_{6,XX}\Lag_{(040)}. \label{eq:LNH6} 
\end{align}
Consequently, the aspect that we need to address is if Eqs. (\ref{eq:LNH2}-\ref{eq:LNH6}) contain any dynamics beyond Eqs. (\ref{eq:LH2}-\ref{eq:LH5}).

Our analysis will be systematic. Starting with the terms with lowest $p$, we will apply all the exact forms and antisymmetric redundancies at that level. Then, we will do the same with the next level. We find that
\begin{enumerate}[(i)] 
\item $p=0$

There are no $\Lag_{i}^{NH}$ at this level. The only possible term of this kind is $\Lag_{(000)}$, which already belongs to $\Lag^{H}_{2}$.

\item $p=1$

At this level we have $\Lag_{2}^{NH}$. This term is very simple because it is proportional to $-2X$, which can be reabsorbed in the free function. More explicitly, using the algebraic relation (\ref{eq:L1}), we can see that 
\begin{equation}
\Lag_{2}^{NH}[E_{2}]=-2X\Lag_{2}^{H}[E_{2}]=\Lag_{2}^{H}[-2XE_{2}].
\end{equation}
Thus, as it was trivial to see, $\Lag_{2}^{NH}$ belongs to $\Lag_{2}^{H}$.
\item $p=2$

Then, we have $\Lag_{3}^{NH}$. Here, we will use the first exact form (\ref{eq:D1}). Additionally, we will apply the next antisymmetric redundancy, (\ref{eq:L2}), to rewrite $\Lag_{(0\bar{1}0)}$ in terms of $\Lag_{(011)}$ and $\Lag_{(010)}$. We obtain that
\begin{equation}
\D\Lag_{(000)}^{D-1}[G_{2}]=G_{2,\phi}\Lag_{(001)}+(G_{2}+2XG_{2,X})\Lag_{(010)}+G_{2,X}\Lag_{(011)}.
\end{equation}
This expression can be rewritten to show that
\begin{equation}
\Lag_{3}^{NH}[G_{2,X}]=-\Lag_{2}^{NH}[G_{2,\phi}]-\Lag_{3}^{H}[(G_{2}+2XG_{2,X})]+\D\Lag_{(000)}^{D-1}[G_{2}].
\end{equation}
Due to the fact that we have already seen that $\Lag_{2}^{NH}\subset\Lag^{H}$, we conclude that $\Lag_{3}^{NH}$ also belongs to Horndeski since the exact form does not modify the e.o.m. One should notice that this structure, in which we obtain that a given $\Lag^{NH}_{i}$ is equal to a combination of $\Lag_{i-1}^{NH}$, $\Lag_{i}^{H}$ and $\D\Lag^{D-1}$, will appear again in the forthcoming cases. Then, we will argue that, since we have proven that $\Lag_{i-1}^{NH}$ is included Horndeski, $\Lag_{i}^{NH}$ also belongs.
\item $p=3$

Now, we have to analyze $\Lag_{3}^{NH}$. We will start with the exact form relation (\ref{eq:D2}). This expression can be rewritten, using the antisymmetric identities (\ref{eq:L3}-\ref{eq:L4}), into 
\begin{equation}
\D\Lag_{(010)}^{D-1}[G_{3}]=G_{3,\phi}\Lag_{(011)}+\frac{1}{2}G_{3,X}\Lag_{(021)}+(G_{3}+XG_{3,X})\Lag_{(020)}+\frac{1}{2}G_{2}(\Lag_{(101)}+2X\Lag_{(100)}).
\end{equation}
Remarkably, the above equation can be translated into
\begin{equation}
\Lag_{4}^{NH}[G_{3}]=-2\Lag_{3}^{NH}[G_{3,\phi}]-2\Lag_{4}^{H}[XG_{3}]+2\D\Lag_{(010)}^{D-1}[G_{3}].
\label{eq:DLNH4}
\end{equation}
Therefore, this result implies that $\Lag_{4}^{NH}$ belongs to Horndeski theory, for same arguments as before and applying that we already know that $\Lag_{3}^{NH}\subset\Lag^{H}$. Interestingly, this result is telling us that $\Lag_{(101)}$ can be seen as a linear combination of Horndeski's theory up to quartic order and $\Lag_{(021)}$. In the discussion, we will extend on this issue.
\item $p=4$

Finally, we have two more Lagrangians: $\Lag_{5}^{NH}$ and $\Lag_{6}^{NH}$. For analyzing $\Lag_{5}^{NH}$, we begin with the exact form (\ref{eq:D3}) to show that 
\begin{equation}
G_{4,X}\Lag_{(1\bar{1}0)}=G_{4,\phi}\Lag_{(101)}+G_{4}\Lag_{(110)}-\D\Lag^{D-1}_{(100)}[G_{4}].
\end{equation}
Then, we use another exact form (\ref{eq:D4}) and apply the algebraic relations (\ref{eq:L5}-\ref{eq:L6})
\begin{equation}
\begin{split}
\D\Lag_{(020)}^{D-1}[G_{4,X}]&=G_{4,\phi}\Lag_{(101)}+G_{4,\phi X}\Lag_{(021)}+(2XG_{4,X}+G_{4})\Lag_{(110)}+(\frac{2X}{3}G_{4,XX}+G_{4,X})\Lag_{030} \\
&+G_{4,X}\Lag_{(111)}+\frac{1}{3}G_{4,XX}\Lag_{031}-\D\Lag^{D-1}_{(100)}[G_{4}],
\end{split}
\end{equation}
which again simplifies into
\begin{equation}
\Lag_{5}^{NH}[G_{4,X}]=-\Lag_{4}^{NH}[G_{4,\phi}]-\Lag_{5}^{H}[2XG_{4,X}+G_{4}]+\D\Lag_{(020)}^{D-1}[G_{4,X}]+\D\Lag^{D-1}_{(100)}[G_{4}].
\label{eq:DLNH5}
\end{equation}
Thus, we find that $\Lag_{5}^{NH}$ belongs to Horndeski theory too, due to the fact that $\Lag_{4}^{NH}\subset\Lag^{H}$. Again, it will be interesting to discuss later the relation between $\Lag_{(111)}$, $\Lag_{(031)}$ and Horndeski theory.

Lastly, we try to uncover $\Lag_{6}^{NH}$. Using the exact form (\ref{eq:D5}) and the antisymmetric relation (\ref{eq:L7}), we realize that we can write $\Lag_{(1\bar{2}0)}$ as
\begin{equation}
G_{5,X}\Lag_{(1\bar{2}0)}=G_{5,\phi}\Lag_{(111)}+G_{5}\Lag_{120}+\frac{X}{2}G_{5}\Lag_{(200)}-\D\Lag^{D-1}_{(110)}[G_{5}].
\end{equation}
Now, plugging the above expression in the last exact form (\ref{eq:D6}) together with the remaining algebraic identities (\ref{eq:L8}) and (\ref{eq:L9}), step by step, we find that
\begin{equation}
\begin{split}
\D\Lag_{(030)}^{D-1}[G_{5,X}]&=G_{5,\phi X}\Lag_{(031)}+3G_{5,\phi}\Lag_{(111)}+(G_{5,X}+\frac{X}{2}G_{5,XX})\Lag_{(040)} \\
&+3(G_{5}+XG_{5,X})\Lag_{(120)}+\frac{3X}{2}G_{5}\Lag_{200}-3\D\Lag^{D-1}_{(110)}[G_{5}],
\end{split}
\end{equation}
implying that
\begin{equation}
\Lag_{6}^{NH}[\frac{3X}{2}G_{5}]=-\Lag_{5}^{NH}[3G_{5,\phi}]+\D\Lag_{(030)}^{D-1}[G_{5,X}]+3\D\Lag^{D-1}_{(110)}[G_{5}].
\end{equation}
Therefore, $\Lag_{6}^{NH}$ also belongs to Horndeski, since we have proven before that $\Lag_{5}^{NH}\subset\Lag^{H}$. This result generalizes the one obtained by \cite{Kobayashi:2011nu}, in which they showed that $f(\phi)GB$ belongs to Horndeski using the e.o.m. Here, we show \emph{explicitly} that a kinetic Gauss-Bonnet term as $\Lag_{6}^{NH}$, which contains the case studied by \cite{Kobayashi:2011nu}, belongs to Horndeski Lagrangian. Furthermore, from the above expression we also learn that if $G_{5}$ is only a function of $X$, $\Lag_{6}^{NH}$ becomes an exact form itself. We should point out that $\Lag_{6}^{NH}$ with a coefficient depending on $\phi$ and $X$ \emph{has not been previously studied} in the literature. Here, we have shown that such a new Lagrangian has second order e.o.m. but its dynamics is already described by Horndeski's theory.
\end{enumerate}

In conclusion, we have seen that there is a total of ten independent Lagrangians, which can be chosen to be the six of Horndeski, plus Beyond Horndeski, i.e. $\Lag_{(021)}$ and $\Lag_{(031)}$, plus $\Lag_{(040)}$ and the Gauss-Bonnet $\Lag_{(200)}$. From them, there are only four independent combinations giving rise to second order e.o.m. These four independent Lagrangians can be chosen to be the ones of Horndeski, i.e. (\ref{eq:LH2}-\ref{eq:LH5}). We realize that with this procedure we are not able to conclude anything whether $\Lag_{(021)}$ and $\Lag_{(031)}$ are well behaved by themselves, as they do in Beyond Horndeski theories ($G^{3}$) \cite{Gleyzes:2014dya}. This would require a Hamiltonian analysis. However, this result tells us that the higher derivative structure of $G^{3}$ model, i.e.  $\Lag_{(021)}$ and $\Lag_{(031)}$, is precisely the same as the one of $\Lag_{(101)}$ and $\Lag_{(111)}$ respectively. This seems to indicate that those terms might also be ghost free.
\section{Discussion}
\label{sec:Discussion}

In this work, we have developed a new formulation for scalar-tensor theories in the language of differential forms. We have found a finite and closed basis that describes general theories of this class in arbitrary dimensions, including Horndeski and the $G^3$ set of Beyond Horndeski theories. Within this basis, we have been able to systematically classify the relations between different physical theories and to find all possible Lagrangians leading to second-order equations of motion in four as well as arbitrary number of spacetime dimensions. At this point, it is crucial to discuss the implications of our work in connection to the recent literature. 

In the field of general scalar-tensor theories, the fundamental analysis was made by Horndeski \cite{Horndeski:1974wa}, who found the most general second order scalar-tensor Euler-Lagrange equations in four dimensions. In practice, Horndeski's theorem was first proven at the level of the equations of motion, imposing a relation between the divergence of the metric and the scalar field equations arising from diffeomorphism invariance. He then classified all the possible terms compatible with this requirement and proceeded by finding an action that produced them in the equations of motion. In this sense, our work has followed the opposite direction. We have started by looking for the most general action satisfying invariance under Local Lorentz Transformations in a pseudo-Riemannian manifold and constructed with a fixed set of building blocks; the vielbein $\theta^{a}$, the curvature 2-form $\R^{ab}$, the 1-form $\Psi^{a}$ encoding first derivatives of the scalar field and the 1-form $\Phi^{a}$ containing second derivatives linearly. Then, we have looked for the combinations which give rise to second-order e.o.m. It is important to remark that in this paper we have not proven Horndeski's theorem, since, as we have discussed in Sec. \ref{sec:RewritingST}, our basis of Lagrangians can be generalized to higher powers of the derivatives of the field, cf. App. \ref{app:HigherOrder}. However, what we \emph{have proven} is that Horndeski's theory corresponds to the most general second order 4-form Lagrangian invariant under LLT in a pseudo-Riemannian manifold and constructed with $\theta^{a}$, $\R^{ab}$, $\Psi^{a}$ and $\Phi^{a}$. Consequently, Horndeski theorem guarantees that any non-linear extension of our basis will be either equivalent to it or characterized by higher derivatives e.o.m.

More recently, Horndeski's theory was encountered in the context of Covariant Galileons \cite{Deffayet:2009wt}. These models are the curved-space extensions of the Galileon theory \cite{Nicolis:2008in} described by second order equations of motion. Galileons in turn represents the most general scalar theory in flat space with only second order derivatives (since the Galileon symmetry $\phi\to\phi+c+b_\mu x^\mu$ eliminates all first derivatives). This model has different formulations that differ only by total derivatives (exact forms in our notation). The original one, denoted by $\Lag_{N}^{Gal,1}$ in \cite{Deffayet:2011gz}, corresponds in our notation to the terms $\Lag_{(0N1)}$, where $N$ runs from 0 to $D-1$. The other relevant formulation, named $\Lag_{N}^{Gal,3}$ also in \cite{Deffayet:2011gz}, can be written in the context of this paper as $X\Lag_{(0N0)}$, where again $N=0, \cdots, D-1$. From these two versions of the Galileon theory, a curved-space extension has been performed. 

On the one hand, Ref. \cite{Deffayet:2009mn} started from $\Lag_{N}^{Gal,1}$. They found a general result in $D$-dimensions that yields second order e.o.m. Their result is equivalent to our Lagrangian (\ref{eq:lagD}) with $n=1$. Since they are considering always terms with $n=1$, they obtain $D$ different Lagrangians, provided that $2l+m+1\leq D$. On the other hand, Ref. \cite{Deffayet:2011gz} found a Covariant Galileon theory from $\Lag_{N}^{Gal,3}$. Their result can be written as our Lagrangian (\ref{eq:lagD}) with $n=0$. Due to the fact that they wanted to reproduce Galileon theory in flat space, they only considered $D$ different Lagrangians. Therefore, they did not considered the possible second order Lagrangian satisfying $2l+m=D$. This Lagrangian $\Lag_{6}^{NH}[E_{6}(\phi,X)]$, which we named \emph{kinetic Gauss-Bonnet} and presented in (\ref{eq:LNH6}), has not been previously studied in the literature. However, we have also shown in this work that its dynamics is already contained in the full Horndeski's theory. A particular case of this Lagrangian, when $E_{6}=E_{6}(\phi)$, is the well-known scalar coupling to the Gauss-Bonnet term $f(\phi)GB$ \cite{Nojiri:2005vv}. In this respect, with the previous result, we have additionally proven \emph{explicitly} that such a theory belongs to Horndeski, as it was claimed in Ref. \cite{Kobayashi:2011nu}. Interestingly, when there is only kinetic dependence in the coefficient, i.e. $E_{6}=E_{6}(X)$, the kinetic Gauss-Bonnet Lagrangian becomes identically an exact form. As an additional remark, with our set of exact forms and algebraic relation, summarized in Fig. \ref{fig:Interconnection}, one can easily link the two covariantized forms of Galileon theory \cite{Deffayet:2011gz,Deffayet:2009mn}. This relation is not trivial, as pointed out by \cite{Deffayet:2011gz}, and it was missing in the literature. 

Lastly, a new line of analysis has been opened for scalar-tensor theories in the last few years. It consists in studying theories with higher than two time derivatives in the e.o.m. but with hidden constraints that save from Ostrogradski's instabilities, implying the existence of viable theories Beyond Horndeski \cite{Zumalacarregui:2013pma}. At the end, the key ingredient to avoid the presence of ghosts is to have a degenerate theory \cite{Langlois:2015cwa}, as it is clearly explained in \cite{Woodard:2015zca}. Lagrangians of this type include the $G^3$ theory \cite{Gleyzes:2014dya,Gleyzes:2014qga}, which in our notation correspond to $\Lag_{(021)}$ and $\Lag_{(031)}$, cf. (\ref{eq:L021}) and (\ref{eq:L031}) respectively. Such theories need a Hamiltonian analysis in order to properly disentangle the physical degrees of freedom \cite{Langlois:2015skt}. In fact, several subtleties can arise to make the theory inviable, such is the case of a generic combination of Horndeski and Beyond Horndeski Lagrangians that becomes non-degenerate even though each term is degenerate by itself \cite{Langlois:2015cwa,Crisostomi:2016tcp}. Within this work, we have focused in scalar-tensor theories with second-order Euler-Lagrange equations and we cannot conclude anything about the viability of this third generation of scalar-tensor theories. However, from our analysis, one learns that the form of these higher derivative terms in the e.o.m. is the same for $\Lag_{(101)}$ and $\Lag_{(021)}$, and for $\Lag_{(111)}$ and $\Lag_{(031)}$. This fact seems to point out that $\Lag_{(101)}$ and $\Lag_{(111)}$, given by (\ref{eq:L101}) and (\ref{eq:L111}), will also propagate only the graviton and the scalar field\footnote{In fact, as pointed out in Sec. \ref{sec:RewritingST}, $\Lag_{(101)}$ and $\Lag_{(111)}$ correspond respectively to ``John" and ``Paul" Lagrangians of the Fab Four theory \cite{Charmousis:2011bf}. In Ref. \cite{Babichev:2015qma}, where extended Fab Four models were studied, it was found the same conclusion, i.e. these Lagrangians can be related to a Beyond Horndeski term plus Horndeski Lagrangians, cf. Eq. (\ref{eq:DLNH4}) and (\ref{eq:DLNH5}).}. As stressed, a specific Hamiltonian analysis in this direction would be needed to confirm this argument.

In the ground of general field theoretical studies of gravity, our work could be seen as a \emph{scalar-tensor extension} of the analysis of Lovelock's theory \cite{Lovelock:1971yv} in differential forms \cite{Mardones:1990qc} (see a recent review in \cite{Charmousis:2008kc}). Moreover, we have been able to \emph{systematically} classify every possible Lagrangian in our basis and present its interconnections with the others, uncovering the internal structure of the scalar-tensor theories. These relations lead to a minimal basis of ten independent Lagrangians, of which four independent combinations produce second-order Euler-Lagrange equations. Along this paper, we have followed the common choice of considering the modern version of Horndeski's theory as the basic set, together with the extra Lagrangians present in $G^3$ Beyond Horndeski theories, plus an additional $\Lag_{(200)}$ and $\Lag_{(040)}$. Ultimately this choice of basis is a matter of taste and we want to emphasize the different possibilities through Fig. \ref{fig:Interconnection}.

As a final remark, it is important to note that our formalism greatly \emph{simplifies} the computations. The fact that the full e.o.m. of a general scalar-tensor theory can be presented in a few lines is an example of the power of this new notation. Then, using the dictionary between differential forms and component notation included in App. \ref{app:Contractions}, the connection with the literature is direct. Moreover, the compact differential form version of the scalar-tensor Lagrangians compared to the usual tensorial form represents a great advantage. Additionally, the well-established relations between different building blocks through derivative operations allow for a simple connection between Lagrangians with a different number of fields and derivatives. For these reasons, the potential application of this new formulation for scalar-tensor theories extends to many possible grounds of research interest.

The simplicity afforded by differential forms encourages a broad range of applications. Our formalism could be easily adapted to study the role of field redefinitions in scalar-tensor theories in a manner analogous to the use of total derivatives and algebraic relations, extending the scope of previous works and potentially finding new sets of equivalences \cite{Zumalacarregui:2012us,Bettoni:2013diz,Bettoni:2015wta,Achour:2016rkg}. These tools are also suited to analyze other phenomenological and theoretical properties of scalar-tensor theories: for example, simplifications of the equations of motion in the presence of symmetries become very transparent in this formalism. Finally, these tools can be applied to a fundamental analysis of the degrees of freedom present in general Lagrangians, paving the way towards the discovery and characterization of the most general scalar-tensor theories of gravity.

\subsection*{Acknowledgements}

We thank David Langlois for beneficial correspondence and comments to the first version of the manuscript. This work is supported by the Research Project of the Spanish MINECO, FPA2013-47986-03-3P, and the Centro de Excelencia Severo Ochoa Program SEV-2012-0249. JME is supported by the FPU grant FPU14/01618. MZ thanks IFT-UAM-CSIC for hospitality during the completion of this work. The computations of the tensorial form of the Lagrangians have been checked using the xAct package for Mathematica \cite{Brizuela:2008ra,xAct}.

\appendix
\section{Notation}
\label{app:Notation}

The aim of this appendix is to summarize the notation used throughout this work. In order to achieve this task, we must firstly review some key concepts about differential geometry. For a complete and detailed introduction to this subject, one could read, for instance, Ref. \cite{nakahara2003geometry}. Afterwards, we will present particular notation of this work introduced to simplify the calculations.

\subsection{Differential Forms}
\label{app:DiffForms}

Let us begin with the fundamental building block in which we rewrite our scalar-tensor theory, a differential $q$-form. A $q$-form is a totally antisymmetric $(0,q)$-tensor. Due to its antisymmetric character, the space of $q$-forms, $\Omega^{q}(\mathcal{M})$, has a finite dimension $\frac{D!}{(D-q)!q!}$, where $D$ is the dimension of the space-time manifold $\mathcal{M}$. For convenience, we will work in a \emph{non-coordinate} basis $\theta^{a}$, assuming, as it is required by physical arguments, that our base manifold $\mathcal M$ posses a metric $\mathfrak{g}$. Then, we can define the metric as
\begin{equation}
\mathfrak{g}=g_{\mu\nu}dx^{\mu}\otimes dx^{\nu}=\eta_{ab}\theta^{a}\otimes\theta^{b},
\label{eq:metric}
\end{equation}
using that $\theta^{a}$ is related with the cotangent basis vector $dx^{\mu}$ through the \emph{vielbeins} $e^{a}_{~\mu}$ by $\theta^{a}=e^{a}_{~\mu}dx^{\mu}$. Here, $\eta_{ab}$ is the Minkowski metric. Along the work, we will use greek indices to represent coordinate components and latin indices for non-coordinate ones.

The basic operations that one can build on the space of $q$-forms $\Omega^{q}(\mathcal{M})$ are:
\begin{enumerate}[(i)] 
\item \emph{Wedge Product}

It is a totally antisymmetric tensor product that maps $\wedge:\Omega^{q}(\mathcal M)\times\Omega^{r}(\mathcal M)\rightarrow\Omega^{q+r}(\mathcal M)$. In components, if we start with $\omega=\frac{1}{q!}\omega_{a_{1}\cdots a_{q}}\theta^{a_{1}}\wedge\cdots\wedge\theta^{a_{q}}$ and $v=\frac{1}{r!}v_{b_{1}\cdots b_{r}}\theta^{b_{1}}\wedge\cdots\wedge\theta^{b_{r}}$, the wedge product of $\omega$ and $v$ is given by
\begin{equation}
\omega\wedge v=\frac{1}{q!r!}\omega_{a_{1}\cdots a_{q}}v_{b_{1}\cdots b_{r}}\theta^{a_{1}}\wedge\cdots\wedge\theta^{a_{q}}\wedge\theta^{b_{1}}\wedge\cdots\wedge\theta^{b_{r}}.
\label{eq:wedge}
\end{equation}
Importantly, this product posses the property $\omega\wedge v=(-1)^{q \cdot r}v \wedge\omega$. Moreover, it can be used to construct the whole space of $q$-forms in $D$-dimensions. In particular, there is only one independent $D$-form, the volume element $\eta$, which can be written as
\begin{equation}
\begin{split}
\eta=&\theta^{1}\wedge\cdots\wedge\theta^{D}=\sqrt{-g} dx^{1}\wedge\cdots\wedge dx^{D},
\end{split}
\label{eq:volume}
\end{equation}
where $g$ corresponds to the determinant of the metric tensor $g_{\mu\nu}$. One should notice that the volume element can be equivalently written as $\eta=\frac{1}{D!}\epsilon_{a_{1}\cdots a_{D}}\theta^{a_{1}}\wedge\cdots\wedge\theta^{a_{D}}$, where $\epsilon_{a_{1}\cdots a_{D}}$ is the totally antisymmetric symbol. The term $\sqrt{-g}$ appears due to the antisymmetrization of the vielbeins when we change from $\theta^{a}$ to $dx^{\mu}$. Naturally, we can now define the \emph{integral} of a function $f$, i.e. a 0-form, over a manifold $\mathcal M$ by $\int_{\mathcal M}f\wedge\eta$.

\item \emph{Exterior Derivative}

It is a derivative operation that maps $d:\Omega^{q}(\mathcal M)\rightarrow\Omega^{q+1}(\mathcal M)$. If we introduce the partial derivative 1-form operator $\partial$, then the exterior derivative is defined by
\begin{equation}
d\omega=\partial\wedge\omega.
\label{eq:extderiv}
\end{equation}
Relevantly, this derivative satisfies a \emph{graded} Leibniz rule $d(\omega\wedge v)=(d\omega)\wedge v+(-1)^{q}\omega\wedge(dv)$. Additionally, it also fulfills that $d^{2}=0$. When a $q$-form $\alpha$ can written in terms of $(q-1)$-form $\beta$ via $\alpha=d\beta$, it is said that $\alpha$ is an \emph{exact form}. Whenever we have a $q$-form $\omega$ such that $d\omega=0$, $\omega$ is called a \emph{closed form}.

\item \emph{Interior Product}

It is an operation that maps $\mathfrak{i}_{X}:\Omega^{q}(\mathcal M)\rightarrow\Omega^{q-1}(\mathcal M)$, where $X$ is a vector field. In components, it contracts the first index of the $q$-form with the vector field, i.e.
\begin{equation}
\mathfrak{i}_{X}\omega=\frac{1}{(q-1)!}X^{a_{1}}\omega_{a_{1}a_{2}\cdots a_{q}}\theta^{a_{2}}\wedge\cdots\wedge\theta^{a_{q}}.
\label{eq:intprod}
\end{equation}
Interestingly, one can now relate the exterior derivative and the Lie derivative through $\mathcal{L}_{X}\omega=\mathfrak{i}_{X}(d\omega)+d(\mathfrak{i}_{X}\omega)$.

\item \emph{Hodge Dual}

It is an operation that arises in manifolds endowed with a metric and maps $\star:\Omega^{q}(\mathcal M)\rightarrow\Omega^{D-q}(\mathcal M)$. Its action on the non-coordinate basis is
\begin{equation}
\star\left(\theta^{a_{1}}\wedge\cdots\wedge\theta^{a_{q}}\right)=\frac{1}{(D-q)!}\epsilon^{a_{1}\cdots a_{q}}_{~~~~~~~a_{p+1}\cdots a_{D}}\theta^{a_{q+1}}\wedge\cdots\wedge\theta^{a_{D}}.
\label{eq:HodgeDual}
\end{equation}
Since, we will need this expression many times, we will dub it the \emph{Hodge dual basis} and denote it by $\theta^{\star}_{~a_{1}\cdots a_{q}}$. In addition, this operation is the dual of the wedge product and it can be used to the define the \emph{inner product} of two $q$-forms $\alpha$ and $\beta$ by $(\alpha,\beta)=\int\alpha\wedge\star\beta$.
\end{enumerate}

\subsection{Differential Geometry}
\label{app:DiffGeom}

Once we have introduced the standard operations, we are ready to present the geometrical quantities characterizing a curved manifold. First, we introduce the \emph{connection 1-form} $\omega^{a}_{~b}$, which is matrix-valued 1-form. Subsequently, we can bring in the \emph{torsion 2-form} $T^{a}$, i.e. $T^{a}=\frac{1}{2}T^{a}_{~bc}\theta^{b}\wedge\theta^{c}$, and the \emph{curvature 2-form} $\mathcal R^{a}_{~b}$, i.e. $\mathcal R^{a}_{~b}=\frac{1}{2}R^{a}_{~bcd}\theta^{c}\wedge\theta^{d}$. The connection is linked to the torsion and curvature through the \emph{Cartan's structure equations},
\begin{align}
&T^{a}=d\theta^{a}+\omega^{a}_{~b}\wedge\theta^{b}, \label{eq:appCartan1} \\
&\mathcal R^{a}_{~b}=d\omega^{a}_{~b}+\omega^{a}_{~c}\wedge\omega^{c}_{~b}. \label{eq:appCartan2}
\end{align}
These equations can be further simplified if we introduce an \emph{exterior covariant derivative} $\D$, constructed from the connection $\omega^{a}_{~b}$. In this notation, we have $T^{a}= \D\theta^{a}$ and $\mathcal R^{a}_{~b}= \D\omega^{a}_{~b}$. Moreover, \emph{Bianchi's identities}, which are just the result of applying $\D$ on Cartan's structure equations, read
\begin{align}
&\D T^{a}=dT^{a}+\omega^{a}_{~b}\wedge T^{b}=\mathcal R^{a}_{~b}\wedge\theta^{b}, \label{eq:appBianchi1} \\
&\D\mathcal R^{a}_{~b}=d\mathcal R^{a}_{~b}+\omega^{a}_{~c}\wedge\mathcal R^{c}_{~b}-\mathcal R^{a}_{~c}\wedge\omega^{c}_{~b}=0. \label{eq:appBianchi2}
\end{align}
Additionally, it will be relevant in the calculations the generalized version of \emph{Stoke's theorem}, 
\begin{equation}
\int_{\mathcal M}\D\omega=\int_{\partial\mathcal M}\omega,
\end{equation}
which summarizes all the usual Calculus integration theorems. With this tool, and assuming that the surface terms vanish, we will be able to eliminate the exact forms from our Lagrangians. 

For the purpose of our physical discussion, we will restrict our analysis to manifolds in which the connection is uniquely determined by the vielbein, i.e. space-times in which the non-metricity and the torsion vanish (see \cite{Ortin:2004ms} for a specific discussion in more general manifolds). These two conditions translate into the antisymmetry of the 1-form connection indices, $\omega_{ab}=-\omega_{ba}$, and into $T^{a}=0$ respectively. In this context, we will be interested in finding the relation between two connections associated to different vielbeins. With this result, we will be able to find the actual relation between $\omega^{ab}$ and $\theta^{a}$. We start with a metric compatible and torsionless connection, i.e. $\omega_{ab}=-\omega_{ba}$, and $T^{a}=0$, associated to a given vielbein $\theta^{a}$. Then, we define another connection $\tilde{\omega}^{ab}$, which is also metric compatible and torsionless, arising from a vielbein $\tilde{\theta}^{a}$. If we parametrize the difference between the two connections with a 1-form $X^{ab}=\tilde{\omega}^{ab}-\omega^{ab}$, the vanishing of the torsion tells us that
\begin{equation}
\tilde{T}^{a}=\tilde{\D}\tilde{\theta}^{a}=\D\tilde{\theta}^{a}+X^{a}_{~b}\wedge\tilde{\theta}^{b}=0.
\end{equation}
Now, using the basic operations of the exterior algebra presented above, we can find a unique solution for $\tilde{\omega}^{ab}$ in terms of $\tilde{\theta}^{a}$ that is also metric compatible, i.e.
\begin{equation}
\tilde{\omega}^{ab}=\omega^{ab}+\frac{1}{2}\lp\mathfrak{i}_{\tilde{e}^{b}}(\D\tilde{\theta}^{a})-\mathfrak{i}_{\tilde{e}^{a}}(\D\tilde{\theta}^{b})+\mathfrak{i}_{\tilde{e}^{a}}(\mathfrak{i}_{\tilde{e}^{b}}(\D\tilde{\theta}_{c}))\tilde{\theta^{c}}\rp,
\label{eq:apptildeW}
\end{equation}
where $\tilde{\theta}^{a}=\tilde{e}^{a}_{~\mu}dx^{\mu}$. This result is the differential form version of the usual tensorial expression for the spin connection used in supergravity, as it can be found, for instance, in Ref. \cite{Ortin:2004ms}. If we impose the \emph{vielbein postulate}, i.e. $\nabla_{\mu}e^{a}_{~\nu}=0$, this result is directly linked to the Levi-Civita connection $\Gamma(g)$. The component expression of $\tilde{\Gamma}(\tilde{g})$ can be found, for example, in \cite{Wald:1984rg}.

In this work, the explicit expression of the 1-form connection (\ref{eq:apptildeW}) will be important because it will allow us to compute the variation of the connection $\delta\omega^{ab}$ as a function of the variation of the frame, given by $\tilde{\theta}^{a}=\theta^{a}+\delta\theta^{a}$. Using that $T^{a}=\D\theta^{a}=0$ and keeping at first order in the perturbations\footnote{A linear perturbation theory implies that if the vielbein is defined as $\tilde{e}^{a}_{~\mu}=e^{a}_{~\mu}+\delta e^{a}_{~\mu}$, its inverse must be $\tilde{e}_{a}^{~\mu}=e_{a}^{~\mu}-\delta e_{a}^{~\mu}$. One should notice also that, at first order, the indices of the perturbed vielbein $\delta e^{a}_{~\mu}$ are raised with the original vielbein $e^{a}_{~\mu}$.}, we find that 
\begin{equation}
\delta\omega^{ab}=\frac{1}{2}\lp\mathfrak{i}_{e^{b}}(\D\delta\theta^{a})-\mathfrak{i}_{e^{a}}(\D\delta\theta^{b})+\mathfrak{i}_{e^{a}}(\mathfrak{i}_{e^{b}}(\D\delta\theta_{c}))\theta^{c}\rp=\nabla^{b}\delta\theta^{a}-\nabla^{a}\delta\theta^{b}.
\end{equation}
This relation will be very useful for computing the vielbein e.o.m. in the second order formalism, see Sec. \ref{subsec:vEOM}.

\subsection{Contractions with the Hodge Dual Basis}
\label{app:Contractions}

Finally, we present the dictionary between the differential forms language used throughout this work and the standard tensorial notation appearing in the literature of scalar-tensor theories. For that purpose, we are going to introduce some extra notation following Ref. \cite{Zumalacarregui:2013pma}. The possible powers of second derivatives of a scalar field $\phi$ are encoded in
\begin{equation}
\left.\Phi^{n}\right._{\mu\nu}=\phi_{;\mu\alpha_{1}}\left.\phi^{;\alpha_{1}}\right._{;\alpha_{2}}\cdots\left.\phi^{;\alpha_{n-1}}\right._{;\nu},
\label{eq:PhiN}
\end{equation}
where the covariant derivatives follow $\nabla_{\mu}\nabla_{\nu}\phi=\phi_{;\mu\nu}$ and $\nabla_{\mu}\phi=\phi_{,\mu}=\partial_{\mu}\phi$. In this context, the Riemann curvature tensor appear via the commutator of two covariant derivatives acting on a vector, i.e. $[\nabla_{\mu},\nabla_{\nu}]v^{\lambda}=R^{\lambda}_{~\gamma\mu\nu}v^{\gamma}$. 

In addition, we denote the contraction of a (0,2)-tensor $t_{\mu\nu}$ with the metric, i.e. the trace, by $[t_{\mu\nu}]\equiv t_{\mu\nu}g^{\mu\nu}$. In the same fashion, we denote its contraction with first derivatives of the scalar field by $\langle t_{\mu\nu}\rangle\equiv \phi^{,\mu}t_{\mu\nu}\phi^{,\nu}$. Applying these concepts to the Riemann tensor, we can have, for instance,
\begin{align}
&[R_{\mu\nu}]=R_{\mu\nu}g^{\mu\nu}, & &\langle R_{\mu\nu}\rangle= \phi^{,\mu}R_{\mu\nu}\phi^{,\nu}, & &\mathrm{and} & &\langle R_{\mu\nu\rho\gamma}\Phi^{\nu\gamma}\rangle= \phi^{,\mu}R_{\mu\nu\rho\gamma}\Phi^{\nu\gamma}\phi^{,\rho}.
\label{eq:egR}
\end{align}
In the case of the contractions of the second derivatives, we can omit the indices inside the brackets since there is no ambiguity, i.e.
\begin{align}
&[\Phi^{n}]=\left.\Phi^{n}\right._{\mu\nu}g^{\mu\nu}, & &\mathrm{and}  & &\langle\Phi^{n}\rangle= \phi^{,\mu}\left.\Phi^{n}\right._{\mu\nu}\phi^{,\nu}.
\label{eq:egR}
\end{align}

Subsequently, we show how to translate a general Lagrangian written in differential forms, such as (\ref{eq:L}), in components. We simply need to use the definitions presented before for the Hodge dual basis and the exterior product. One should notice that, as a Lagrangian consists in a $D$-form, it is going to be proportional to the volume element $\eta$, since there is only one independent $D$-form. Then, the component structure of the total set of wedge products can be read from $\theta^{a_{1}}\wedge\cdots\wedge\theta^{a_{D}}=\epsilon^{a_{1}\cdots a_{D}}\eta$. Afterwards, the remaining free indices can be contracted using the definition of the Hodge dual basis in (\ref{eq:HodgeDual}). Lastly, we only need to recall the component expression for the different building blocks of the theory, i.e. $\R^{a}_{~b}=\frac{1}{2}R^{a}_{~bcd}\theta^{c}\wedge\theta^{d}$, $\Phi^{a}=\nabla^{a}\nabla_{b}\phi\theta^{b}$ and $\Psi^{a}=\nabla^{a}\phi\nabla_{b}\phi\theta^{b}$. Therefore, a general Lagrangian given by (\ref{eq:L}) can be written in components as
\begin{equation}
\Lag_{(lmn)}=\frac{\eta}{2^{l}(D-p)!}\prod_{i=0}^{l}R^{a_{i}b_{i}}_{~~~~e_{i}f_{i}}\prod_{j=0}^{m}\phi^{;c_{j}}_{~~;g_{j}}\prod_{k=0}^{n}\phi^{,d_{k}}\phi_{,h_{k}}\epsilon_{a_{1}b_{1}\cdots a_{l}b_{l}c_{1}\cdots c_{m}d_{1}\cdots d_{n}p_{1}\cdots p_{D-N}}\epsilon^{e_{1}f_{1}\cdots e_{l}f_{l}g_{1}\cdots g_{m}h_{1}\cdots h_{n}p_{1}\cdots p_{D-p}},
\end{equation}
where $p=2l+m+n$. To exemplify this general recipe, we can particularize for specific cases, for instance,
\begin{align}
\Lag_{(010)}&=\Phi^{a}\wedge\theta^{\star}_{~a}=\frac{1}{3!}\phi^{;a}_{~~;e}\epsilon_{abcd}\epsilon^{ebcd}\eta=[\Phi]\eta, \\
\Lag_{(100)}&=\R^{ab}\wedge\theta^{\star}_{~ab}=\frac{1}{2\cdot2!}R^{ab}_{~~ef}\epsilon_{abcd}\epsilon^{efcd}\eta=R\eta, \\
\Lag_{(030)}&=\Phi^{a}\wedge\Phi^{b}\wedge\Phi^{c}\wedge\theta^{\star}_{~abc}=\phi^{;a}_{~~;e}\phi^{;b}_{~~;f}\phi^{;c}_{~~;g}\epsilon_{abcd}\epsilon^{efgd}\eta=([\Phi]^3 -3[\Phi][\Phi^2]+2[\Phi^3])\eta.
\end{align}

\section{Generalizations}
\label{app:MostGeneral}
\subsection{Higher Order Lagrangians}
\label{app:HigherOrder}

In this appendix, we are going to present a generalization of the basis of Lagrangians $\Lag_{(lmn)}$ introduced in (\ref{eq:L}). We are going to consider building blocks for our theory that depend non-linearly on the power of derivatives of the fields. In particular, we are going to substitute our 1-form encoding the second derivatives of the scalar $\Phi^{a}$, given in (\ref{eq:2deriv}), by 
\begin{equation}
\left(\Phi^{n}\right)^{a}\equiv\left.\Phi^{n}\right.^{a}_{~b}\theta^{b},
\end{equation}
which contains any possible contraction of the field's second derivatives. Also, we are going to generalize the first derivative 1-form $\Psi^{a}$, defined in (\ref{eq:1deriv}), to
\begin{equation}
\left(\Psi^{mn}\right)^{a}\equiv\left.\Phi^{m}\right.^{a}_{~b}\phi^{,b}\phi^{,c}\left.\Phi^{n}\right._{cd}\theta^{d},
\end{equation}
where $\left.\Phi^{n}\right.^{a}_{~b}$ can be found in (\ref{eq:PhiN}). With these new building blocks and imposing invariance under Local Lorentz Transformations, we can construct a generalized version of $\Lag_{(lmn)}$ (\ref{eq:L}) as
\begin{equation}
\left.\mathcal{L}^{(uv)}\right._{(lm_{1}\cdots m_{u}n_{1}\cdots n_{st})}=\bigwedge_{i=1}^{l}\mathcal{R}^{a_{i}b_{i}}\wedge\bigwedge_{r=1}^{u}\bigwedge_{j_{r}=1}^{m_{r}}\left(\Phi^{r}\right)^{c_{j_{r}}}\wedge\bigwedge_{s=1}^{v}\bigwedge_{t=1}^{s}\bigwedge_{k_{st}=1}^{n_{st}}\left(\Psi^{st}\right)^{d_{k_{st}}}\wedge\theta^{\star}_{~a_{1}b_{1}\cdots a_{l}b_{l}c_{1}\cdots c_{m_{r}}\cdots c_{m_{u}}d_{1}\cdots d_{n_{st}}\cdots d_{n_{vv}}},
\label{eq:Lrst}
\end{equation}
which shares the same structure of (\ref{eq:L}) but including any possible higher order 1-form $\left(\Phi^{n}\right)^{a}$ and $\left(\Psi^{mn}\right)^{a}$. The difference is that, now, for maximum power $u$, we have $u$ possible building blocks $\left(\Phi^{m}\right)^{a}$ appearing $m_{u}$ times each. Also, for maximum power $v$, we have $v^{2}$ possible building blocks $\left(\Psi^{mn}\right)^{a}$ appearing $n_{st}$ times each, where $s,t<v$.

In this more general framework, we can accommodate the extended basis $\Lag_{(\bar{l}mn)}$ and $\Lag_{(l\bar{m}n)}$ presented in (\ref{eq:Lbarm}) and (\ref{eq:Lbarl}), which were formed contracting with partial derivatives of the scalar field, introducing $\left(\Psi^{01}\right)^{a}=\phi^{,a}\phi^{,b}\left.\Phi\right._{bc}\theta^{c}$. The only terms that are not contained are those involving a direct contraction of the curvature 2-form with gradients of the scalar field.
\subsection{Pontryagin Forms}
\label{app:PontryaginForms}

Here, we are going to show the terms that complete the set of Lagrangians $\Lag_{(lmn)}$ to give the most general basis satisfying invariance under Local Lorentz Transformations in a pseudo-Riemannian manifold and constructed with the vielbein $\theta^{a}$, the curvature 2-form $\R^{ab}$, $\Psi^{a}$ and $\Phi^{a}$. However, as we are going to argue, they are not very interesting because they cannot give rise to second order e.o.m.

These extra terms appear by direct contraction of the indices of the building blocks. They are the scalar-tensor equivalent of the \emph{Pontryagin forms} in Lovelock-Cartan theories \cite{Mardones:1990qc}. Since we cannot introduce the Hodge dual basis $\theta^{\star}_{a_{1}\cdots a_{k}}$, they must satisfy that $p\equiv2l+m+n=D$. In $4D$, we obtain that there are five possible terms, labeled with an upper $P$ from Pontryagin, but only three are no zero, i.e.
\begin{align}
& \Lag^{P}_{1}= \R^{ab}\wedge \R_{ab}, \label{eq:P1} \\
& \Lag^{P}_{2}=\R^{ab}\wedge \Phi_{a} \wedge \Phi_{b}, \label{eq:P2} \\
& \Lag^{P}_{3}=\R^{ab}\wedge \Phi_{a} \wedge \Psi_{b}, \label{eq:P3} \\
& \Lag^{P}_{4}=\Phi^{a} \wedge \Phi^{b}\wedge \Phi_{a} \wedge \Phi_{b}=0, \label{eq:P4} \\
& \Lag^{P}_{5}=\Phi^{a} \wedge \Phi^{b}\wedge \Phi_{a} \wedge \Psi_{b}=0, \label{eq:P5}
\end{align}
where we have used in the last two lines that $\Phi^{a}\wedge\Phi_{a}=\nabla^{a}\nabla_{b}\phi\nabla_{a}\nabla_{c}\theta^{b}\wedge\theta^{c}=\left.\Phi^{2}\right._{bc}\theta^{b}\wedge\theta^{c}=0$, which is a consequence of the symmetry of the indices of $\left.\Phi^{m}\right._{ab}$. As a comment, one could notice that $\Lag^{P}_{1}= \R^{ab}\wedge \R_{ab}$ is a topological term, since it does not depend on the vielbein. In fact, it is the only topological term, apart from the Gauss-Bonnet $\Lag_{(200)}=\R^{ab}\wedge \R^{cd}\wedge\theta^{\star}_{abcd}$, characterizing a pseudo-Riemannian manifold in $4D$.

Due to the fact that we have computed the Euler-Lagrange equations for a general $\Lag_{(lmn)}$, we can easily analyze the case of Pontryagin forms. As we have extensively discussed in Sec. \ref{sec:EOM}, there are higher than two derivatives terms associated with the variation of each Lagrangian that must be canceled, in order to avoid Ostrogradski instabilities. The case under study now is similar to case studied in which $p=4$, where we found two viable combinations $\Lag_{5}^{NH}$ and $\Lag_{6}^{NH}$, given in (\ref{eq:LNH5}) and (\ref{eq:LNH6}). However, there is an important difference now. Two of the Lagrangians with $p=4$ are identically zero, i.e. (\ref{eq:P4}) and (\ref{eq:P5}). Consequently, they cannot be used to erase the higher derivative of the other terms. In conclusion, there cannot be constructed Lagrangians with second order e.o.m. out of the Pontryagin forms.

\section{Explicit Computations}
\label{app:Computations}

\subsection{Contracted Lagrangians in 4D}
\label{app:Contracted}

In this appendix, we present the explicit component expression for the contracted Lagrangians arising from $\Lag_{(\bar{l}mn)}$, (\ref{eq:Lbarl}), and $\Lag_{(l\bar{m}n)}$, (\ref{eq:Lbarm}), in $4D$. We find

\begin{enumerate}[(i)] \itemsep0pt \parskip0pt \parsep0pt
\item $p=1$
\begin{align}
\mathcal{L}_{(0\bar{1}0)}&=\phi_{,a}\Phi^{a}\wedge\theta^{\star}_{~b}\phi^{,b}=\langle\Phi\rangle\eta \\
\mathcal{L}_{(00\bar{1})}&=\phi_{,a}\Psi^{a}\wedge\theta^{\star}_{~b}\phi^{,b}=(-2X)\Psi^{a}\wedge\theta^{\star}_{~a}=4X^{2}\eta
\intertext{\item $p=2$}
\mathcal{L}_{(\bar{1}00)}&=\phi_{,a}\mathcal{R}^{ab}\wedge\theta^{\star}_{~cb}\phi^{,c}=\langle R_{ab}\rangle\eta \\
\mathcal{L}_{(0\bar{2}0)}&=\phi_{,a}\Phi^{a}\wedge\Phi^{b}\wedge\theta^{\star}_{~cb}\phi^{,c}=(\langle\Phi\rangle[\Phi] -\langle\Phi^2\rangle)\eta \\
\mathcal{L}_{(0\bar{1}1)}&=\phi_{,a}\Phi^{a}\wedge\Psi^{b}\wedge\theta^{\star}_{~cb}\phi^{,c}=0 \\
\mathcal{L}_{(01\bar{1})}&=\phi_{,b}\Phi^{a}\wedge\Psi^{b}\wedge\theta^{\star}_{~ca}\phi^{,c}=(-2X)\Phi^{a}\wedge\Psi^{b}\wedge\theta^{\star}_{~ab}
\intertext{\item $p=3$}
\mathcal{L}_{(\bar{1}10)}&=\phi_{,a}\mathcal{R}^{ab}\wedge\Phi^{c}\wedge\theta^{\star}_{~dbc}\phi^{,d}=(\langle R_{ab}\rangle[\Phi]-\langle R_{ab}\Phi^{bc}\rangle-\langle R_{abcd}\Phi^{bd}\rangle)\eta \\
\mathcal{L}_{(1\bar{1}0)}&=\phi_{,c}\mathcal{R}^{ab}\wedge\Phi^{c}\wedge\theta^{\star}_{~dab}\phi^{,d}=(R\langle \Phi\rangle-2\langle R_{ab}\Phi^{bc}\rangle)\eta \\
\mathcal{L}_{(\bar{1}01)}&=\phi_{,a}\mathcal{R}^{ab}\wedge\Psi^{c}\wedge\theta^{\star}_{~dbc}\phi^{,d}=0 \\
\mathcal{L}_{(10\bar{1})}&=\phi_{,c}\mathcal{R}^{ab}\wedge\Psi^{c}\wedge\theta^{\star}_{~dab}\phi^{,d}=(-2X)\mathcal{R}^{ab}\wedge\Psi^{c}\wedge\theta^{\star}_{~abc} \\
\mathcal{L}_{(0\bar{3}0)}&=\phi_{,a}\Phi^{a}\wedge\Phi^{b}\wedge\Phi^{c}\wedge\theta^{\star}_{~dbc}\phi^{,d}=(2\langle\Phi^3\rangle-2\langle\Phi^{2}\rangle[\Phi]+\langle \Phi\rangle([\Phi]^{2}-[\Phi^2]))\eta \\
\mathcal{L}_{(0\bar{2}1)}&=\phi_{,a}\Phi^{a}\wedge\Phi^{b}\wedge\Psi^{c}\wedge\theta^{\star}_{~dbc}\phi^{,d}=0 \\
\mathcal{L}_{(02\bar{1})}&=\phi_{,c}\Phi^{a}\wedge\Phi^{b}\wedge\Psi^{c}\wedge\theta^{\star}_{~dab}\phi^{,d}=(-2X)\Phi^{a}\wedge\Phi^{b}\wedge\Psi^{c}\wedge\theta^{\star}_{~abc}
\intertext{\item $p=4$}
\mathcal{L}_{(\bar{2}00)}&=\phi_{,a}\mathcal{R}^{ab}\wedge\mathcal{R}^{cd}\wedge\theta^{\star}_{~ebcd}\phi^{,e}=(\langle R_{abcd}R^{ebcd}\rangle-2\langle R_{ab}R^{bc}\rangle-2\langle R_{abcd}R^{bd}\rangle+\langle R_{ab}\rangle R)\eta \\
\mathcal{L}_{(\bar{1}20)}&=\phi_{,a}\mathcal{R}^{ab}\wedge\Phi^{c}\wedge\Phi^{d}\wedge\theta^{\star}_{~ebcd}\phi^{,e} \\
 &=2(\langle R_{ab}\rangle([\Phi]^{2}-[\Phi^2])-2[\Phi](\langle R_{ab}\Phi^{bc}\rangle+\langle R_{abcd}\Phi^{bd}\rangle)+2(\langle R^{ab}\left.\Phi^{2}\right._{bc}\rangle+\langle R^{abcd}\left.\Phi^{2}\right._{bd}\rangle+\langle R^{abcd}\Phi_{bd}\Phi_{ec}\rangle))\eta \nonumber\\
\mathcal{L}_{(1\bar{2}0)}&=\phi_{,c}\mathcal{R}^{ab}\wedge\Phi^{c}\wedge\Phi^{d}\wedge\theta^{\star}_{~eabd}\phi^{,e} \\
&=(R(\langle\Phi\rangle[\Phi]-\langle\Phi^2\rangle)-2(\langle R_{ab}\Phi^{bc}\rangle[\Phi]+\langle \Phi\rangle[R\Phi])+2(\langle R^{ab}\left.\Phi^{2}\right._{bc}\rangle+\langle \Phi_{ab}R^{bc}\Phi_{cd}\rangle+\langle R^{abcd}\Phi_{bd}\Phi_{ec}\rangle))\eta \nonumber\\
\mathcal{L}_{(\bar{1}11)}&=\phi_{,a}\mathcal{R}^{ab}\wedge\Phi^{c}\wedge\Psi^{d}\wedge\theta^{\star}_{~ebcd}\phi^{,e}=0 \\
\mathcal{L}_{(1\bar{1}1)}&=\phi_{,c}\mathcal{R}^{ab}\wedge\Phi^{c}\wedge\Psi^{d}\wedge\theta^{\star}_{~eabd}\phi^{,e}=0 \\
\mathcal{L}_{(11\bar{1})}&=\phi_{,d}\mathcal{R}^{ab}\wedge\Phi^{c}\wedge\Psi^{d}\wedge\theta^{\star}_{~eabc}\phi^{,e}=(-2X)\mathcal{R}^{ab}\wedge\Phi^{c}\wedge\Psi^{d}\wedge\theta^{\star}_{~abcd} \\
\mathcal{L}_{(0\bar{4}0)}&=\phi_{,a}\Phi^{a}\wedge\Phi^{b}\wedge\Phi^{c}\wedge\Phi^{d}\wedge\theta^{\star}_{~ebcd}\phi^{,e} \\
&=(6\langle\Phi^4\rangle -6\langle\Phi^3\rangle[\Phi]+3\langle\Phi^2\rangle[\Phi]^2-\langle\Phi\rangle[\Phi]^{3}-3\langle\Phi^{2}\rangle[\Phi^{2}]+3\langle\Phi\rangle[\Phi][\Phi^{2}]-2\langle\Phi\rangle[\Phi^{3}])\eta \nonumber\\
\mathcal{L}_{(0\bar{3}1)}&=\phi_{,a}\Phi^{a}\wedge\Phi^{b}\wedge\Phi^{c}\wedge\Psi^{d}\wedge\theta^{\star}_{~ebcd}\phi^{,e}=0 \\
\mathcal{L}_{(03\bar{1})}&=\phi_{,d}\Phi^{a}\wedge\Phi^{b}\wedge\Phi^{c}\wedge\Psi^{d}\wedge\theta^{\star}_{~abce}\phi^{,e}=(-2X)\Phi^{a}\wedge\Phi^{b}\wedge\Phi^{c}\wedge\Psi^{d}\wedge\theta^{\star}_{~abcd}
\end{align}
\end{enumerate}
As it is clear in the above results, any $(\bar{l}m1)$ or $(l\bar{m}1)$ term is zero since we are contracting two equal vectors with an antisymmetric tensor (it is the same argument that limits $n$ to be 0 or 1). In addition, the terms of the form $(lm\bar{1})$ do not introduce new structures since they are equal to $(lm1)$ with an extra $(-2X)$ factor in front. In total, there are only 10 independent Lagrangians.

\subsection{Scalar Equations of Motion in 4D}
\label{app:sEOM}

We continue the calculation presented in Sec. \ref{sec:EOM}, for the scalar e.o.m.:

\begin{enumerate}
\item[(iv)] $p=3$

When $n=0$, we have two different contributions, $G_{5}\Lag_{(110)}=G_{5}\wedge\R^{ab}\wedge\Phi^{c}\wedge\theta^{\star}_{~abc}$ and $F_{5}\Lag_{(030)}=F_{5}\wedge\Phi^{a}\wedge\Phi^{b}\wedge\Phi^{c}\wedge\theta^{\star}_{~abc}$. We proceed as before and analyze each term separately. First, we have
\begin{equation}
\begin{split}
\delta(G_{5}\Lag_{(110)})=&\delta G_{5}\wedge\R^{ab}\wedge\Phi^{c}\wedge\theta^{\star}_{~abc}+G_{5}\wedge\R^{ab}\wedge\delta\Phi^{c}\wedge\theta^{\star}_{~abc} \\
=&\delta\phi\wedge\left(G_{5,\phi}\wedge\R^{ab}\wedge\Phi^{c}\wedge\theta^{\star}_{~abc}-\nabla^{z}\left(G_{5,X}\nabla_{z}\phi\right)\wedge\R^{ab}\wedge\Phi^{c}\wedge\theta^{\star}_{~abc}\right) \\
+&\delta\phi\wedge G_{5,X}\nabla_{z}\phi\wedge\left(\uline{\nabla^{z}\R^{ab}}\wedge\Phi^{c}+\R^{ab}\wedge\dashuline{\nabla^{z}\Phi^{c}}\right)\wedge\theta^{\star}_{~abc} \\
+&\delta\phi\wedge\left(\nabla^{c}\left(G_{5,\phi}\D\phi\right)\wedge\R^{ab}\wedge\theta^{\star}_{~abc}-\nabla^{c}\left(G_{5,X}\nabla_{z}\phi\right)\wedge\Phi^{\alpha}\wedge\R^{ab}\wedge\theta^{\star}_{~abc}\right) \\
+&\delta\phi\wedge\left(-G_{5,X}\nabla_{z}\phi\wedge\dashuline{\nabla^{c}\Phi^{z}}\wedge\R^{ab}\wedge\theta^{\star}_{~abc}+\D\left(G_{5}\right)\wedge\nabla^{c}\R^{ab}\wedge\theta^{\star}_{~abc}\right)+\mathcal{O}\left(G_{5,[\Phi]}\right).
\end{split}
\label{eq:H3a}
\end{equation}
Using the commutation of covariant derivatives (\ref{eq:id1}), we can eliminate the unwanted contribution of the higher derivative terms, underlined with a dashed line. Also we can see that the second term of the last line is identically zero due to Bianchi's identity and the antisymmetry of its indices. The only remaining higher order term is underlined with a solid line.

Then, we have
\begin{equation}
\begin{split}
\delta(F_{5}\Lag_{(030)}&)=\delta F_{5}\wedge\Phi^{a}\wedge\Phi^{b}\wedge\Phi^{c}\wedge\theta^{\star}_{~abc}+3F_{5}\wedge\delta\Phi^{a}\wedge\Phi^{b}\wedge\Phi^{c}\wedge\theta^{\star}_{~abc} \\
=&\delta\phi\wedge\left(F_{5,\phi}\wedge\Phi^{a}\wedge\Phi^{b}\wedge\Phi^{c}\wedge\theta^{\star}_{~abc}+\nabla^{z}\left(F_{5,X}\nabla_{z}\phi\right)\wedge\Phi^{a}\wedge\Phi^{b}\wedge\Phi^{c}\wedge\theta^{\star}_{~abc}\right) \\
+&\delta\phi\wedge\left(3F_{5,X}\nabla_{z}\phi\wedge\dashuline{\nabla^{z}\Phi^{a}}+3\nabla^{a}\left(F_{5,\phi}\D\phi\right)\right)\wedge\Phi^{b}\wedge\Phi^{c}\wedge\theta^{\star}_{~abc} \\
-3&\delta\phi\wedge\left(\nabla^{a}\left(F_{5,X}\nabla_{z}\phi\right)\wedge\Phi^{z}\wedge\Phi^{b}+F_{5,X}\nabla_{z}\phi\wedge\left(\dashuline{\nabla^{a}\Phi^{z}}\wedge\Phi^{b}+2\Phi^{z}\wedge\nabla^{a}\Phi^{b}\right)\right)\wedge\Phi^{c}\wedge\theta^{\star}_{~abc} \\
+6&\delta\phi\wedge\left(\nabla^{a}\left(F_{5}\nabla_{z}\phi\right)\wedge\R^{bz}\wedge\Phi^{c}+F_{5}\nabla_{z}\phi\wedge\left(\uline{\nabla^{a}\R^{bz}}\wedge\Phi^{c}+\R^{bz}\wedge\nabla^{a}\Phi^{c}\right)\right)\wedge\theta^{\star}_{~abc}+\mathcal{O}\left(F_{5,[\Phi]}\right).
\end{split}
\label{eq:H3b}
\end{equation}
Doing a similar analysis as before, we can see that the dashed underlined terms add up in the appropriate way. Moreover, the third terms of both the fourth and last line are well behaved due to the antisymmetry of their indices. Finally, we are left again with only one higher order contribution, i.e. the solid underlined term. Remarkably we can eliminate the higher terms of (\ref{eq:H3a}) and (\ref{eq:H3b}) by applying the Bianchi identity (\ref{eq:id2}). We obtain that this cancellation occurs if $F_{5}=\frac{1}{3}G_{5,X}$. As in all the previous cases, we also must to impose $G_{5}=G_{5}(\phi,X)$ to avoid higher than 2nd order terms in the e.o.m. Doing so, we obtain the Lagrangian presented in Eq. (\ref{eq:H5}).

When $n=1$, we also have two different contributions, $E_{4}\Lag_{(101)}=E_{4}\wedge\R^{ab}\wedge\Psi^{c}\wedge\theta^{\star}_{~abc}$ and $H_{4}\Lag_{(021)}=H_{4}\wedge\Phi^{a}\wedge\Phi^{b}\wedge\Psi^{c}\wedge\theta^{\star}_{~abc}$. We continue similarly to previous cases and investigate term by term. First, we have
\begin{equation}
\begin{split}
\delta(E_{4}\Lag_{(101)})=&\delta E_{4}\wedge\R^{ab}\wedge\Psi^{c}\wedge\theta^{\star}_{~abc}+E_{4}\wedge\R^{ab}\wedge\delta\Psi^{c}\wedge\theta^{\star}_{~abc} \\
=&\delta\phi\wedge\left(E_{4,\phi}\wedge\R^{ab}\wedge\Psi^{c}\wedge\theta^{\star}_{~abc}+\nabla^{z}\left(E_{4,X}\nabla_{z}\phi\wedge\Psi^{c}\right)\wedge\R^{ab}\wedge\theta^{\star}_{~abc}\right) \\
+&\delta\phi\wedge E_{4,X}\nabla_{z}\phi\wedge\uline{\nabla^{z}\R^{ab}}\wedge\Psi^{c}\wedge\theta^{\star}_{~abc} \\
-&\delta\phi\wedge\left(\nabla^{a}\left(E_{4}\D\phi\wedge\R^{bc}\right)+\D\left(E_{4}\nabla^{a}\phi\wedge\R^{bc}\right)\right)\wedge\theta^{\star}_{~abc}+\mathcal{O}\left(E_{4,[\Phi]}\right).
\end{split}
\label{eq:NH3a}
\end{equation}
In this expression, the only higher order term is underlined with a solid line. Then, we have
\begin{equation}
\begin{split}
\delta(H_{4}\Lag_{(021)})=&\delta H_{4}\wedge\Phi^{a}\wedge\Phi^{b}\wedge\Psi^{c}\wedge\theta^{\star}_{~abc}+2H_{4}\wedge\delta\Phi^{a}\wedge\Phi^{b}\wedge\Psi^{c}\wedge\theta^{\star}_{~abc}+H_{4}\wedge\Phi^{a}\wedge\Phi^{b}\wedge\delta\Psi^{c}\wedge\theta^{\star}_{~abc} \\
=&\delta\phi\wedge\left(H_{4,\phi}\wedge\Phi^{a}\wedge\Phi^{b}\wedge\Psi^{c}\wedge\theta^{\star}_{~abc}+\nabla^{z}\left(H_{4,X}\nabla_{z}\phi\wedge\Psi^{c}\right)\wedge\Phi^{a}\wedge\Phi^{b}\wedge\theta^{\star}_{~abc}\right) \\
+2&\delta\phi\wedge\nabla^{a}\left(H_{4,\phi}\D\phi\wedge\Phi^{b}\wedge\Psi^{c}-H_{4}\wedge\Phi^{b}\wedge \D\left(\Psi^{c}\right)\right)\wedge\theta^{\star}_{~abc} \\
+2&\delta\phi\wedge\left(H_{4,X}\nabla_{z}\phi\wedge\dashuline{\nabla^{z}\Phi^{a}}\wedge\Phi^{b}\wedge\Psi^{c}-\nabla^{a}\left(H_{4,X}\nabla_{z}\phi\wedge\Psi^{c}\right)\wedge\Phi^{z}\wedge\Phi^{b}\right)\wedge\theta^{\star}_{~abc} \\
-2&\delta\phi\wedge H_{4,X}\nabla_{z}\phi\wedge\left(\dashuline{\nabla^{a}\Phi^{z}}\wedge\Phi^{b}+\Phi^{\alpha}\wedge\nabla^{a}\Phi^{b}\right)\wedge\Psi^{c}\wedge\theta^{\star}_{~abc} \\
+2&\delta\phi\wedge\left(\nabla^{a}\left(H_{4}\nabla_{z}\phi\wedge\Psi^{c}\right)\wedge\R^{bz}+H_{4}\nabla_{z}\phi\wedge\uline{\nabla^{a}\R^{bz}}\wedge\Psi^{c}\right)\wedge\theta^{\star}_{~abc} \\
-&\delta\phi\wedge\left(\nabla^{a}\left(H_{4}\D\phi\wedge\Phi^{b}\wedge\Phi^{c}\right)+\D\left(H_{4}\nabla^{a}\phi\wedge\Phi^{b}\wedge\Phi^{c}\right)\right)\wedge\theta^{\star}_{~abc}+\mathcal{O}\left(H_{4,[\Phi]}\right).
\end{split}
\label{eq:NH3b}
\end{equation}
Similarly to previous cases, after applying the the commutator of covariant derivatives (\ref{eq:id1}), we are left with only one higher order contribution. We can eliminate the higher terms of (\ref{eq:NH3a}) and (\ref{eq:NH3b}) by applying the Bianchi identity (\ref{eq:id2}). We obtain that this cancellation occurs if $H_{4}=E_{4,X}$. As in all the previous cases, we also must to impose $E_{4}=E_{4}(\phi,X)$ to avoid higher than 2nd order terms in the e.o.m. Thus, we obtain the Lagrangian presented in Eq. (\ref{eq:NH4}).

\item[(v)] $p=4$

In this last case, we find five different Lagrangians: $E_{5}\Lag_{(111)}=E_{5}\wedge\R^{ab}\wedge\Phi^{c}\wedge\Psi^{d}\wedge\theta^{\star}_{~abcd}$, $H_{5}\Lag_{(031)}=H_{5}\wedge\Phi^{a}\wedge\Phi^{b}\wedge\Phi^{c}\wedge\Psi^{d}\wedge\theta^{\star}_{~abcd}$, $E_{6}\Lag_{(200)}=E_{6}\wedge\R^{ab}\wedge\R^{cd}\wedge\theta^{\star}_{~abcd}$, $E_{7}\Lag_{(120)}=E_{7}\wedge\R^{ab}\wedge\Phi^{c}\wedge\Phi^{d}\wedge\theta^{\star}_{~abcd}$ and $H_{6}\Lag_{(040)}=H_{6}\wedge\Phi^{a}\wedge\Phi^{b}\wedge\Phi^{c}\wedge\Phi^{d}\wedge\theta^{\star}_{~abcd}$. As usual, we investigate them one by one.

The first Lagrangian yields
\begin{equation}
\begin{split}
\delta(E_{5}\Lag_{(111)})=&\lp\delta E_{5}\wedge\R^{ab}\wedge\Phi^{c}\wedge\Psi^{d}+E_{5}\wedge\R^{ab}\wedge\delta\Phi^{c}\wedge\Psi^{d}+E_{5}\wedge\R^{ab}\wedge\Phi^{c}\wedge\delta\Psi^{d}\rp\wedge\theta^{\star}_{~abcd} \\
=&\delta\phi\wedge\left(E_{5,\phi}+\nabla^{z}\left(E_{5,X}\nabla_{z}\phi\right)\right)\wedge\R^{ab}\wedge\Phi^{c}\wedge\Psi^{d}\wedge\theta^{\star}_{~abcd} \\
+&\delta\phi\wedge E_{5,X}\nabla_{z}\phi\wedge\left(\uline{\nabla^{z}\R^{ab}}\wedge\Phi^{c}\wedge\Psi^{d}+\R^{ab}\wedge\dashuline{\nabla^{z}\Phi^{c}}\wedge\Psi^{d}+\R^{ab}\wedge\Phi^{c}\wedge\nabla^{z}\Psi^{d}\right)\wedge\theta^{\star}_{~abcd} \\
+&\delta\phi\wedge\lp\nabla^{c}\left(E_{5,\phi}\D\phi\wedge\Psi^{d}+E_{5}\Phi^{d}\wedge \D\phi\right)+\nabla^{c}\left(E_{5,X}\nabla_{z}\phi\wedge\Psi^{d}\right)\wedge\Phi^{z}\rp\wedge\R^{ab}\wedge\theta^{\star}_{~abcd} \\
+&\delta\phi\wedge\left(-E_{5,X}\nabla_{z}\phi\wedge\dashuline{\nabla^{c}\Phi^{z}}\wedge\R^{ab}\wedge\Psi^{d}+\D\left(E_{5}\wedge\Psi^{d}\right)\wedge\nabla^{c}\R^{ab}\right)\wedge\theta^{\star}_{~abcd} \\
+&\delta\phi\wedge\left(\nabla^{d}\left(E_{5}\D\phi\wedge\R^{ab}\wedge\Phi^{c}\right)+\D\left(E_{5}\nabla^{d}\phi\wedge\R^{ab}\wedge\Phi^{c}\right)\right)\wedge\theta^{\star}_{~abcd}+\mathcal{O}\left(E_{5,[\Phi]}\right).
\end{split}
\label{eq:NH4c}
\end{equation}

The second one gives
\begin{equation}
\begin{split}
\delta(H_{5}\Lag_{(031)})=&\delta H_{5}\wedge\Phi^{a}\wedge\Phi^{b}\wedge\Phi^{c}\wedge\Psi^{d}\wedge\theta^{\star}_{~abcd}+H_{5}\wedge\Phi^{a}\wedge\Phi^{b}\wedge\left(3\delta\Phi^{c}\wedge\Psi^{d}+\Phi^{c}\wedge\delta\Psi^{d}\right)\wedge\theta^{\star}_{~abcd} \\
=&-\delta\phi\wedge\left(H_{5,\phi}\wedge\Psi^{d}+\nabla^{z}\left(H_{5,X}\nabla_{z}\phi\wedge\Psi^{d}\right)\right)\wedge\Phi^{a}\wedge\Phi^{b}\wedge\Phi^{c}\wedge\theta^{\star}_{~abcd} \\
+3&\delta\phi\wedge\left(H_{5,X}\nabla_{z}\phi\wedge\dashuline{\nabla^{z}\Phi^{a}}\wedge\Psi^{d}+\nabla^{a}\left(H_{5,X}\nabla_{z}\phi\wedge\Psi^{d}\right)\wedge\Phi^{z}\right)\wedge\Phi^{b}\wedge\Phi^{c}\wedge\theta^{\star}_{~abcd} \\
+3&\delta\phi\wedge\nabla^{a}\left(H_{5,\phi}\D\phi\wedge\Psi^{d}+H_{5}\wedge \D\left(\Psi^{d}\right)\right)\wedge\Phi^{b}\wedge\Phi^{c}\wedge\theta^{\star}_{~abcd} \\
-3&\delta\phi\wedge H_{5,X}\nabla_{z}\phi\wedge\left(\dashuline{\nabla^{a}\Phi^{z}}\wedge\Phi^{b}+2\Phi^{z}\wedge\nabla^{a}\Phi^{b}\right)\wedge\Phi^{c}\wedge\Psi^{d}\wedge\theta^{\star}_{~abcd} \\
-6&\delta\phi\wedge\nabla^{a}\left(H_{5}\nabla_{z}\phi\wedge\Psi^{d}\right)\wedge\R^{bz}\wedge\Phi^{c}\wedge\theta^{\star}_{~abcd} \\
+6&\delta\phi\wedge\left(H_{5}\nabla_{z}\phi\wedge\left(\uline{\nabla^{a}\R^{bz}}\wedge\Phi^{c}+\R^{bz}\wedge\nabla^{a}\Phi^{c}\right)\wedge\Psi^{d}\wedge\theta^{\star}_{~abcd}\right) \\
+&\delta\phi\wedge(\nabla^{d}(H_{5}\D\phi\wedge\Phi^{a}\wedge\Phi^{b}\wedge\Phi^{c})+\D(H_{5}\nabla^{d}\phi\wedge\Phi^{a}\wedge\Phi^{b}\wedge\Phi^{c}))\wedge\theta^{\star}_{~abcd}+\mathcal{O}\left(H_{5,[\Phi]}\right).
\end{split}
\label{eq:NH4e}
\end{equation}
Applying the same analysis as in previous cases, and setting $H_{5}=\frac{1}{3}E_{5,X}$, we obtain a Lagrangian with second order e.o.m. This Lagrangian was already presented in (\ref{eq:NH5}).

Continuing with the e.o.m., we analyze the third term. It corresponds to the Gauss-Bonnet Lagrangian, which only depends on $\phi$ through its coefficient $E_{6}$. Thus, we find
\begin{equation}
\begin{split}
\delta(E_{6}\Lag_{(200)})=&\delta E_{6}\wedge\R^{ab}\wedge\R^{cd}\wedge\theta^{\star}_{~abcd} \\
=&\delta\phi\wedge\left(\left(E_{6,\phi}+\nabla^{z}\left(E_{6,X}\nabla_{z}\phi\right)\right)\wedge\R^{ab}+2E_{6,X}\nabla_{z}\phi\wedge\uline{\nabla^{z}\R^{ab}}\right)\wedge\R^{cd}\wedge\theta^{\star}_{~abcd}+\mathcal{O}\left(E_{6,[\Phi]}\right).
\end{split}
\label{eq:NH4a}
\end{equation}

The fourth one will be very similar to (\ref{eq:H3b}). We obtain
\begin{equation}
\begin{split}
\delta(E_{7}\Lag_{(120)})&=\delta E_{7}\wedge\R^{ab}\wedge\Phi^{c}\wedge\Phi^{d}\wedge\theta^{\star}_{~abcd}+2E_{7}\wedge\R^{ab}\wedge\delta\Phi^{c}\wedge\Phi^{d}\wedge\theta^{\star}_{~abcd} \\
=&\delta\phi\wedge\left(E_{7,\phi}+\nabla^{\alpha}\left(E_{7,X}\nabla_{\alpha}\phi\right)\right)\wedge\R^{ab}\wedge\Phi^{c}\wedge\Phi^{d}\wedge\theta^{\star}_{~abcd} \\
+&\delta\phi\wedge\lp E_{7,X}\nabla_{z}\phi\wedge\left(\uline{\nabla^{z}\R^{ab}}\wedge\Phi^{c}+2\R^{ab}\wedge\dashuline{\nabla^{z}\Phi^{c}}\right)\wedge\Phi^{d}+2\nabla^{c}\left(E_{7}\R^{ab}\nabla_{z}\phi\right)\wedge\R^{dz}\rp\wedge\theta^{\star}_{~abcd} \\
+2&\delta\phi\wedge\left(\nabla^{c}\left(E_{7,\phi}\D\phi\wedge\R^{ab}\wedge\Phi^{d}\right)+\nabla^{c}\left(E_{7,X}\nabla_{z}\phi\wedge\R^{ab}\wedge\Phi^{d}\right)\wedge\Phi^{z}\right)\wedge\theta^{\star}_{~abcd} \\
-2&\delta\phi\wedge\left(E_{7,X}\nabla_{z}\phi\wedge\R^{ab}\wedge\dashuline{\nabla^{c}\Phi^{z}}\wedge\Phi^{d}-E_{7}\nabla_{z}\phi\uline{\nabla^{a}\R^{bz}}\wedge\R^{cd}\right)\wedge\theta^{\star}_{~abcd} +\mathcal{O}\left(E_{7,[\Phi]}\right) \\
\end{split}
\label{eq:NH4b}
\end{equation}

Finally, the last possibility becomes
\begin{equation}
\begin{split}
\delta(H_{6}\Lag_{(040)})&=\delta H_{6}\wedge\Phi^{a}\wedge\Phi^{b}\wedge\Phi^{c}\wedge\Phi^{d}\wedge\theta^{\star}_{~abcd}+4H_{6}\wedge\delta\Phi^{a}\wedge\Phi^{b}\wedge\Phi^{c}\wedge\Phi^{d}\wedge\theta^{\star}_{~abcd} \\
=&\delta\phi\wedge\left(\left(H_{6,\phi}+\nabla^{z}\left(H_{6,X}\nabla_{z}\phi\right)\right)\wedge\Phi^{a}+4H_{6,X}\nabla_{z}\phi\wedge\dashuline{\nabla^{z}\Phi^{a}}\right)\wedge\Phi^{b}\wedge\Phi^{c}\wedge\Phi^{d}\wedge\theta^{\star}_{~abcd} \\
+4&\delta\phi\wedge\left(\nabla^{a}\left(H_{6,\phi}\D\phi\wedge\Phi^{b}\wedge\Phi^{c}\wedge\Phi^{d}\right)-\nabla^{a}\left(H_{6,X}\nabla_{z}\phi\right)\wedge\Phi^{z}\wedge\Phi^{b}\wedge\Phi^{c}\wedge\Phi^{d}\right)\wedge\theta^{\star}_{~abcd} \\
-4&\delta\phi\wedge\left(H_{6,X}\nabla_{z}\phi\wedge\left(\dashuline{\nabla^{a}\Phi^{z}}\wedge\Phi^{b}+3\Phi^{z}\wedge\nabla^{a}\Phi^{b}\right)-3\nabla^{a}\left(H_{6}\nabla_{z}\phi\right)\wedge\R^{bz}\rp\wedge\Phi^{c}\wedge\Phi^{d}\wedge\theta^{\star}_{~abcd} \\
+12&\delta\phi\wedge H_{6}\nabla_{\alpha}\phi\wedge\left(\uline{\nabla^{a}\R^{bz}}\wedge\Phi^{c}+2\R^{bz}\wedge\nabla^{a}\Phi^{c}\rp\wedge\Phi^{d}\wedge\theta^{\star}_{~abcd}+\mathcal{O}\left(H_{6,[\Phi]}\right).
\end{split}
\label{eq:NH4d}
\end{equation}
Then, again, in order to cancel the higher derivatives underlined terms we must choose $H_{6}=\frac{1}{6}E_{7,X}$ and $E_{7}=2E_{6,X}$. In that case, we obtain the Lagrangian (\ref{eq:NH6}).
\end{enumerate}
\subsection{Exact Forms in 4D}
\label{app:ExactForms}

Here, we present the detailed computation of the exact forms in $4D$. We classify the possible terms as before, depending on the number of fields, $p=2l+m+n$. We obtain:
\begin{enumerate}[(i)] 
\item $p=0$
\begin{align}
\D\mathcal{L}_{(000)}^{D-1}=&\D(G_{2}\wedge\theta^{\star}_{~a}\nabla^{a}\phi) \\
=&\left(G_{2,\phi}\wedge \D\phi-G_{2,X}\nabla_{\alpha}\phi\wedge\Phi^{\alpha}\right)\wedge\theta^{\star}_{~a}\nabla^{a}\phi+G_{2}\wedge\Phi^{a}\wedge\theta^{\star}_{~a} \nonumber\\
=&G_{2,\phi}\wedge \Psi^{a}\wedge\theta^{\star}_{~a}-G_{2,X}\nabla_{\alpha}\phi\wedge\Phi^{\alpha}\wedge\theta^{\star}_{~a}\nabla^{a}\phi+G_{2}\wedge\Phi^{a}\wedge\theta^{\star}_{~a} \nonumber\\
=&G_{2,\phi}\Lag_{(001)}-G_{2,X}\Lag_{(0\bar{1}0)}+G_{2}\Lag_{(010)}, \nonumber
\intertext{\item $p=1$}
\D\mathcal{L}_{(010)}^{D-1}=&\D(G_{3}\wedge\Phi^{a}\wedge\theta^{\star}_{~ba}\nabla^{b}\phi) \\
=&G_{3,\phi}\wedge \Psi^{b}\wedge\Phi^{a}\wedge\theta^{\star}_{~ba}-G_{3,X}\nabla_{\alpha}\phi\wedge\Phi^{\alpha}\wedge\Phi^{a}\wedge\theta^{\star}_{~ba}\nabla^{b}\phi \nonumber\\
+&G_{3}\wedge\mathcal{R}^{a\alpha}\nabla_{\alpha}\phi\wedge\theta^{\star}_{~ba}\nabla^{b}\phi-G_{3}\Phi^{a}\wedge\Phi^{b}\wedge\theta^{\star}_{~ba} \nonumber\\
=&G_{3,\phi}\Lag_{(011)}-G_{3,X}\Lag_{(0\bar{2}0)}+G_{3}\left(\Lag_{(020)}-\Lag_{(\bar{1}00)}\right) \nonumber
\intertext{\item $p=2$}
\D\mathcal{L}_{(100)}^{D-1}=&\D(G_{4}\wedge\mathcal{R}^{ab}\wedge\theta^{\star}_{~abc}\nabla^{c}\phi) \\
=&G_{4,\phi}\wedge \Psi^{c}\wedge\mathcal{R}^{ab}\wedge\theta^{\star}_{~abc}-G_{4,X}\nabla_{\alpha}\phi\wedge\Phi^{\alpha}\wedge\mathcal{R}^{ab}\wedge\theta^{\star}_{~abc}\nabla^{c}\phi+G_{4}\wedge\mathcal{R}^{ab}\wedge\Phi^{c}\wedge\theta^{\star}_{~abc} \nonumber\\
=&G_{4,\phi}\Lag_{(101)}-G_{4,X}\Lag_{(1\bar{1}0)}+G_{4}\Lag_{(110)} \nonumber\\
\nonumber\\
\D\mathcal{L}_{(020)}^{D-1}=&\D(F_{4}\wedge\Phi^{a}\wedge\Phi^{b}\wedge\theta^{\star}_{~abc}\nabla^{c}\phi) \\
=&F_{4,\phi}\wedge \Psi^{c}\wedge\Phi^{a}\wedge\Phi^{b}\wedge\theta^{\star}_{~abc}-F_{4,X}\nabla_{\alpha}\phi\wedge\Phi^{\alpha}\wedge\Phi^{a}\wedge\Phi^{b}\wedge\theta^{\star}_{~abc}\nabla^{c}\phi \nonumber\\
+2&F_{4}\wedge\mathcal{R}^{a\alpha}\nabla_{\alpha}\phi\wedge\Phi^{b}\wedge\theta^{\star}_{~abc}\nabla^{c}\phi+F_{4}\Phi^{a}\wedge\Phi^{b}\wedge\Phi^{c}\wedge\theta^{\star}_{~abc} \nonumber\\
=&F_{4,\phi}\Lag_{(021)}-F_{4,X}\Lag_{(0\bar{3}0)}+F_{4}\left(\Lag_{(030)}-2\Lag_{(\bar{1}10)}\right) \nonumber
\intertext{\item $p=3$}
\D\mathcal{L}_{(110)}^{D-1}=&\D(G_{5}\wedge\mathcal{R}^{ab}\wedge\Phi^{c}\wedge\theta^{\star}_{~dabc}\nabla^{d}\phi) \\
=&G_{5,\phi}\wedge \Psi^{d}\wedge\mathcal{R}^{ab}\wedge\Phi^{c}\wedge\theta^{\star}_{~dabc}-G_{5,X}\nabla_{\alpha}\phi\wedge\Phi^{\alpha}\wedge\mathcal{R}^{ab}\wedge\Phi^{c}\wedge\theta^{\star}_{~dabc}\nabla^{d}\phi \nonumber\\
+&G_{5}\wedge\mathcal{R}^{ab}\wedge\mathcal{R}^{c\alpha}\nabla_{\alpha}\phi\wedge\theta^{\star}_{~dabc}\nabla^{d}\phi-G_{5}\wedge\mathcal{R}^{ab}\wedge\Phi^{c}\wedge\Phi^{d}\wedge\theta^{\star}_{~dabc} \nonumber\\
=&G_{5,\phi}\Lag_{(111)}-G_{5,X}\Lag_{(1\bar{2}0)}+G_{5}\left(\Lag_{(120)}-\Lag_{(\bar{2}00)}\right) \nonumber\\
\nonumber\\
\D\mathcal{L}_{(030)}^{D-1}=&\D(F_{5}\wedge\Phi^{a}\wedge\Phi^{b}\wedge\Phi^{c}\wedge\theta^{\star}_{~abcd}\nabla^{d}\phi) \\
=&F_{5,\phi}\wedge \Psi^{d}\wedge\Phi^{a}\wedge\Phi^{b}\wedge\Phi^{c}\wedge\theta^{\star}_{~abcd}-F_{5,X}\nabla_{\alpha}\phi\wedge\Phi^{\alpha}\wedge\Phi^{a}\wedge\Phi^{b}\wedge\Phi^{c}\wedge\theta^{\star}_{~abcd}\nabla^{d}\phi \nonumber\\
+3&F_{5}\wedge\mathcal{R}^{a\alpha}\nabla_{\alpha}\phi\wedge\Phi^{b}\wedge\Phi^{c}\wedge\theta^{\star}_{~abcd}\nabla^{d}\phi-F_{5}\wedge\Phi^{a}\wedge\Phi^{b}\wedge\Phi^{c}\wedge\Phi^{d}\wedge\theta^{\star}_{~abcd} \nonumber\\
=&F_{5,\phi}\Lag_{(031)}-F_{5,X}\Lag_{(0\bar{4}0)}+F_{5}\left(\Lag_{(040)}-3\Lag_{(\bar{1}20)}\right) \nonumber
\end{align}
\end{enumerate}
Altogether, the above expressions represent six constraints in the basis of scalar-tensor Lagrangians. A graphical representation of them can be found in Fig. \ref{fig:Interconnection}.

\bibliographystyle{h-physrev}
\bibliography{STDiffFormsBib}

\end{document}